\documentclass[epj-spec]{svjour}
\usepackage{graphics}
\usepackage{epsfig,color}
\begin{document}
\title{Modulational instability in isolated and driven Fermi-Pasta-Ulam lattices}
\author{Thierry Dauxois\inst{1}\fnmsep\thanks{\email{Thierry.Dauxois@ens-lyon.fr}}
\and Ramaz Khomeriki\inst{2,3}\fnmsep\thanks{\email{khomeriki@hotmail.com}}
\and Stefano Ruffo\inst{2}\fnmsep\thanks{\email{Stefano.Ruffo@unifi.it}}   }
\institute{Laboratoire de Physique, UMR-CNRS 5672, ENS Lyon, 46
All\'{e}e d'Italie, 69364 Lyon c\'{e}dex 07, France \and
Dipartimento di Energetica, ``S. Stecco'' and CSDC, Universit{\`a}
di  Firenze, and INFN, via S. Marta, 3, 50139 Firenze, Italy \and
Department of Exact and Natural Sciences, Tbilisi State University,
3 Chavchavadze avenue, Tbilisi 0128, Georgia}

\abstract{We present a detailed analysis of the modulational
instability of the zone-boundary mode for one and higher-dimensional
Fermi-Pasta-Ulam (FPU) lattices. The growth of the instability is followed by 
a process of relaxation to equipartition, which we have
called the {\em Anti-FPU problem} because the energy is initially
fed into the highest frequency part of the spectrum, while in the
original FPU problem low frequency excitations of the
lattice were considered. This relaxation process leads to the formation of {\em chaotic
breathers} in both one and two space dimensions. The system then relaxes to
energy equipartition, on time scales that increase as the energy
density is decreased. We supplement this study by considering the 
nonconservative case, where the FPU lattice is homogeneously driven at
high frequencies. Standing and travelling nonlinear waves and
solitonic patterns are detected in this case. Finally we investigate 
the dynamics of the FPU chain when one end is driven at a frequency located 
above the zone boundary. We show that this excitation stimulates 
nonlinear bandgap transmission effects.}
\maketitle
\section{Introduction}
\label{intro}

In 1955, reporting about one of the first numerical simulations,
Fermi, Pasta and Ulam (FPU)~\cite{FPU} remarked that it was {\em $\ldots$
very hard to observe the rate of ``thermalization'' or mixing $\ldots$}
in a nonlinear one-dimensional lattice in which the energy was
initially fed into the lowest frequency mode.  Even if the
understanding of this problem has advanced significantly
afterwards~\cite{ford,lichtlieb}, several issues are still far from
being clarified. In most cases, the evolution towards energy {\em equipartition} 
among linear modes has been checked considering an
initial condition where all the energy of the system was concentrated
in a small packet of modes centered around some low frequency.

Beginning with the pioneering paper of Zabusky and Deem~\cite{ZabuskyDeem},
the opposite case in which the energy is put into a high frequency
mode has been also analyzed. In this early paper, the zone--boundary mode
was excited with an added spatial modulation for the one-dimensional
$\alpha$-FPU model (quadratic nonlinearity in the equations of motion).
Here, we will study the time-evolution of this mode without any
spatial modulation for the $\beta$-FPU model (cubic nonlinearity in the
equations of motion) and some higher--order nonlinearities. Moreover,
we will extend the study to higher dimensional lattices. Since the energy
is fed into the opposite side of the linear spectrum, we call this problem
the {\em Anti-FPU problem}.

In a paper by Bundinsky and Bountis~\cite{bountis}, the
zone--boundary mode solution of the one-dimensional FPU lattice
was found to be unstable above an energy threshold $E_c$ which
scales like $1/N$, where $N$ is the number of oscillators. This result
was later and independently confirmed by Flach~\cite{flach} and Poggi
et al~\cite{PR}, who also obtained the correct factor in the large
$N$-limit. These results were obtained by a direct linear stability
analysis around the periodic orbit corresponding to the zone-boundary
mode. Similar methods have been recently applied to other modes and
other FPU-like potentials by Chechin et al~\cite{chechin,chechin2004}
and Rink~\cite{Rink2003}.

A formula for $E_c$, valid for all $N$, has been obtained in
Refs.~\cite{sandusky,burlakov,DRT,Paladin} in the rotating wave
approximation, and will be also discussed  in this paper.  Associated
with this instability is the calculation of the growth rates of mode
amplitudes. The appropriate approach for Klein-Gordon lattices was first
introduced by Kivshar and Peyrard~\cite{KivsharPeyrard}, following an
analogy with the Benjamin-Feir instability in fluid mechanics~\cite{benjaminFeir}.

Previously, a completly different approach to describe this
instability was introduced by Zakharov and Shabat~\cite{zakharov},
studying the associated Nonlinear Schr{\"o}dinger equation in the
continuum limit. A value for the energy threshold was obtained in
Ref.~\cite{lukomskii} in the continuum limit. The full derivation
starting from the FPU equation of motions was then independently
obtained by Berman and Kolovskii~\cite{berman} in the so-called
``narrow-packet'' approximation.

Only very recently the study of what happens after the modulational
instability develops has been performed for Klein-Gordon~\cite{daumont} and
FPU-lattices\cite{maledetirusi,sandusky}. From these analyses it
turned out that these high-frequency initial conditions lead to a
completely new dynamical behavior in the transient time preceeding
the final energy equipartition.  In particular, the main discovery
has been the presence on the lattice of sharp localized
modes~\cite{maledetirusi,daumont}.  These latter papers were the
first  to make the connection between energy relaxation and
intrinsic localized modes~\cite{ST}, or
breathers~\cite{reviewbreather}. Later on, a careful numerical and
theoretical study of the dynamics of a $\beta$-FPU model was
performed~\cite{cretegny}. It has been shown that moving breathers
play a relevant role in the transient dynamics and that, contrary to
exact breathers, which are periodic solutions, these have a chaotic
evolution. This is why they have been called {\em chaotic
breathers}. Following these studies, Lepri and
Kosevich~\cite{kosevichlepri} and Lichtenberg and
coworkers~\cite{ulman,mirnov} have further characterized the scaling
laws of relaxation times using continuum limit equations.

Zabusky et al. have recently simulated numerically
the behavior of the one-dimensional, periodic $\alpha$-FPU model with
optical and acoustic initial excitations of small-but finite and
large amplitudes. Using beautiful color representations~\cite{zabuskysun} 
of the numerical results, they
find nearly recurrent solutions, where the optical result is due to
the appearance of localized breather-like packets. For large
amplitudes, they obtained also chaotic behaviors for the alpha
lattice.

Using a theory (originally developed~\cite{kimthierry,kim} for the discrete nonlinear
Schr\"odinger equation) where standard Gibbsian equilibrium
statistical mechanics was considered to predict macroscopic average
values for a thermalized state in the thermodynamic limit, Johansson
has recently analyzed~\cite{magnus} certain aspects of a mixed
Klein-Gordon/FPU chain. In particular, he shows that the available
phase space is divided into two separated parts with qualitatively
different properties in thermal equilibrium: one part corresponding
to a normal thermalized state with exponentially small probabilities
for large-amplitude excitations, and another part typically
associated with the formation of high-amplitude localized
excitations, interacting with an infinite-temperature phonon bath.
Observing the $\beta$-FPU chains in the thermalized state,
Gershgorin et al showed~\cite{gershgorin} via numerical simulation
that discrete breathers actually persist and have a turbulent-like
behavior. They describe the dynamical scenario as spatially highly
localized discrete breathers riding chaotically on spatially
extended, renormalized waves.

Recently Flach et al~\cite{flachqbreathers} have
focused on the main FPU observation that the initially excited
normal mode shares its energy for long times only with a few other
modes from a frequency neighbourhood in modal space. They have
identified this long lasting regime as a dynamical localization
effect and applied the methods developed for discrete breathers in
FPU chains to the dynamics of normal modes. The result is that
time-periodic and modal-space-localized orbits, (they call
q-breathers) persist in the FPU model. The dynamics generated by one
initially excited mode evolves close to the related q-breathers for
very long times. Thus many features of the short- and medium-time
evolution of natural packets are encoded in the profile of these
objects.

Let us briefly comment about the chaoticity of these
spontaneously created breathers. Tailleur and Kurchan have recently
implemented~\cite{tailleurkurchan} an efficient method that allows
one to select trajectories with unusual chaoticity, with Lyapunov
weighted dynamics (LWD) (a method originally proposed in the context
of chemical reactions~\cite{tanasakurchan}). As an example of
application, they study the Fermi-Pasta-Ulam nonlinear chain
starting from a microcanonical equilibrium configuration. They show
that the algorithm rapidly singles out the chaotic-breathers when
searching for trajectories with high level of chaoticity (typically
they study cases where the Lyapunov is three times the one of a
typical equilibrium trajectory), thus confirming that the large
Lyapunov configurations are dominated by chaotic breathers.

Most of the previous studies are for one-dimensional
lattices. We have recently derived modulational instability
thresholds also for higher dimensional lattices~\cite{chaos2005} and
we have presented a study of chaotic breathers formation in
two-dimensional FPU lattices. However, the study is extremely
preliminary and further analyses are needed. In particular, the full
process of relaxation to energy equipartion and the associated time
scales have not been carefully studied in two-dimensional FPU lattices.
Pioneering results on the relaxation process in a two-dimensional
triangular FPU lattice from low frequency initial states seem to
indicate a faster evolution to equipartition~\cite{benettin}.
Benettin has discovered that for large values of the energy per site
the time scale for equipartition can be quite short, even in the
thermodynamic limit of the lattice size, and that this time scale
increases as only one-over the energy per site. If one lowers the
energy per site below a critical threshold, however, the time scale
for equipartition on finite lattices grows much more rapidly. But
the critical threshold value of energy per site appears to vanish as
the lattice size goes to infinity. A similar analysis for high
frequencies remains to be performed.

Further, having already studied the process of formation of stable
localized  structures arising from modulational instability in the
conservative case~\cite{cretegny}, we are strongly motivated to see
how the presence of forcing and damping affects this process. To
remain close to the  Hamiltonian case, we restrict ourselves to the
case of small damping. Various types of forcing are in principle
possible, depending on the physical situation under study. However,
a general requirement for localization is to excite band-edge modes.
For Klein-Gordon lattices  this is naturally realized using a
spatially uniform driving field, which has been shown to induce
interesting pattern formation phenomena~\cite{burlakov2}. On the
other hand, this forcing would not be effective for FPU lattices,
because, due to the symmetry of the Hamiltonian, the zero mode is
decoupled. Alternatively, since spatial localization appears from
the instability of band-edge modes, we choose in the context of Anti
FPU scenario to drive the system near the zone boundary
wavelength. As it will be shown below stationary localized patterns
(either moving or static) appear under such a homogeneous driving
and damping.

Finally, we consider driving the system by one end, again with 
frequencies above the zone boundary. By this we stimulate the
appearance of a ``supratransmission" scenario \cite{leon} in FPU, i.e.
the chain becomes conductive only for driving amplitudes above
a threshold \cite{supra}. This phenomenon is explained in terms of
nonlinear response manifolds in Ref.~\cite{Maniadis}.

We have organized the paper in the following way.  In
Section~\ref{modinst}, the modulational instability of zone-boundary
modes on the lattice is discussed, beginning with the
one-dimensional case, followed by the two-dimensional and higher
dimensional cases and finishing with the continuum nonlinear
Schr{\"o}dinger approach.  In this Section, we also decribe the
mechanism of creation of chaotic breathers in one and two
dimensions. Section~\ref{localization1d} deals with the driven-damped
Anti-FPU scenario (homogeneous and point-like driving). Some final
remarks and conclusions are reported in Section~\ref{conclusions}.

\section{Modulational Instability}
\label{modinst}

\subsection{The one-dimensional case}
\label{modinst1D}

We will discuss in this section modulational instability for the
one-dimensional FPU lattice, where the linear coupling is corrected
by a $(2\ell+1)$th order nonlinearity, with $\ell$ a positive
integer. Denoting by $u_n(t)$ the relative displacement of the
$n$-th particle from its equilibrium position, the equations of
motion are
\begin{eqnarray}
\ddot{u}_n = u_{n+1} + u_{n-1} - 2u_n + (u_{n+1}-u_n)^{2\ell+1} -
(u_n - u_{n-1})^{2\ell+1}. \label{sub}
\end{eqnarray}
We adopt a lattice of $N$ particles and we choose periodic boundary
conditions. For the sake of simplicity, we first report on  the
analysis for $\ell=1$ (i.e. for the  $\beta$-FPU model) and then we
generalize the results to any $\ell$-value.

Due to periodic boundary conditions, the normal modes
associated to the linear part of Eq.~(\ref{sub})
are plane waves of the form
\begin{equation}
u_n(t) =\frac{a}{2}\left(e^{i\theta_n (t)}+e^{-i\theta_n (t)}\right)
\label{plane}
\end{equation}
where $\theta_n(t) = qn-\omega t$ and $q=2\pi k/N$ ($k=-N/2,\dots ,N/2$).  The
dispersion relation of nonlinear phonons in the rotating wave
approximation~\cite{sandusky} is $\omega^2(q)=4(1+\alpha)\sin ^{2} (q/2)$, where $\alpha=3a^2
\sin^2(q/2)$ takes into account the nonlinearity.  Modulational
instability of such a plane wave is investigated by studying the
linearized equation associated with the envelope of the carrier
wave~(\ref{plane}).  Therefore, one introduces infinitesimal
perturbations in the amplitude and phase and looks for solutions of
the form
\begin{eqnarray}
u_n(t) &=& \frac{a}{2}[ 1 + b_n (t) ]\ e^{i\left[\theta_n (t)+\psi_n(t)\right]}+
 \frac{a}{2}[ 1 + b_n (t) ]\ e^{-i\left[\theta_n (t)+\psi_n(t)\right]}\nonumber\\
&=&a[1+b_n(t)]\ \cos[qn-\omega t+\psi_n(t)],
\label{az}
\end{eqnarray}
where  $b_n$ and $\psi_n$ are reals and assumed to be small
in comparison with the parameters of the carrier wave.
Substituting Eq.~(\ref{az}) into the equations of motion and
keeping the second derivative,
we obtain for the real and imaginary part of the secular term
$e^{i(qn-\omega t)}$ the following equations
\begin{eqnarray}
-\omega^2b_n+2\omega\dot\psi_n+\ddot b_n&=&(1+2\alpha)
\left[\cos q\,(b_{n+1}+ b_{n-1})-2b_n \right]\nonumber \\
&-&\alpha\left(b_{n+1}+ b_{n-1}-2b_n\cos q \right)
-(1+2\alpha)\sin q\,(\psi_{n+1}-\psi_{n-1}) \label{eq4} \\
-\omega^2\psi_n-2\omega\dot b_n+\ddot \psi_n&=&
(1+2\alpha)\left[\cos q\, (\psi_{n+1}+\psi_{n-1})-2\psi_n\right]
\nonumber \\
&+&(1+2\alpha)\sin q\,(b_{n+1}-b_{n-1})+\alpha
\left(\psi_{n+1}+ \psi_{n-1}-2\psi_n\cos q  \right).\label{eq5}
\end{eqnarray}

Further assuming  $b_n=b_0 \ e^{i(Qn-\Omega t)}+{\rm c.c.}$
and $\psi_n=\psi_0 \ e^{i(Qn-\Omega t)}+{\rm c.c.}$
we obtain  the following equations for the secular term
$e^{i(Qn-\Omega t)}$
\begin{eqnarray}
b_0\Bigl[\Omega^2+\omega^2+2(1+2\alpha)(\cos q \cos Q -1)
&-&2\alpha(\cos Q-\cos q) \Bigr] \nonumber \\
&-&2i\psi_0\left[\omega\Omega+(1+2\alpha)\sin q\sin Q \right]=0\label{eq6}\\
\psi_0\Bigl[ \Omega^2+\omega^2+2(1+2\alpha)(\cos q\cos Q-1)
&+&2\alpha(\cos Q-\cos q)\Bigr] \nonumber \\
&+&2ib_0\left[ \omega\Omega +(1+2\alpha)\sin q\sin Q\right]=0.\label{eq7}
\end{eqnarray}
In the case of Klein-Gordon type
equations~\cite{KivsharPeyrard,daumont}, one neglects the second order
derivatives in Eqs.~(\ref{eq4})-(\ref{eq5}). This can be justified by
the existence of a gap in the dispersion relation for $q=0$, which
allows to neglect $\Omega^2$ with respect to $\omega^2$. In the FPU case, this
approximation is worse, especially for long wavelengths, because there
is no gap.

Non trivial solutions for Eqs.~(\ref{eq6})-(\ref{eq7}) can be
found only if the Cramer's determinant vanishes, i.e. if the
following equation is fulfilled:
\begin{eqnarray}
\Biggl[(\Omega+\omega)^2-4(1+2\alpha)\sin^2\left({q+Q\over2}\right)\Biggr]&&
\Biggl[(\Omega-\omega)^2-
4(1+2\alpha)\sin^2\left({q-Q\over2}\right)\Biggr]  \nonumber \\
&=& 4\alpha^2\left(\cos{Q}-\cos q\right)^2.
\label{relatdispercorr}
\end{eqnarray}

This equation admits four different solutions when the wavevectors $q$
of the unperturbed wave and $Q$ of the perturbation are fixed.  If one
of the solutions is complex, an instability of one of the modes $(q\pm
Q)$ is present, with a growth rate equal to the imaginary part of the
solution. Using this method, one can derive the instability threshold
amplitude for any wavenumber. A trivial example is the case of $q=0$,
for which we obtain $\Omega=\pm \sin\left({Q/ 2}\right)$, which proves that
the zero mode solution is stable. This mode is present due to the
invariance of the equations of motion~(\ref{sub}) with respect to the
translation $u_n \to u_n+const$ and, as expected, is completely
decoupled from the others.

\begin{figure}[ht]
\centerline{\epsfig{file=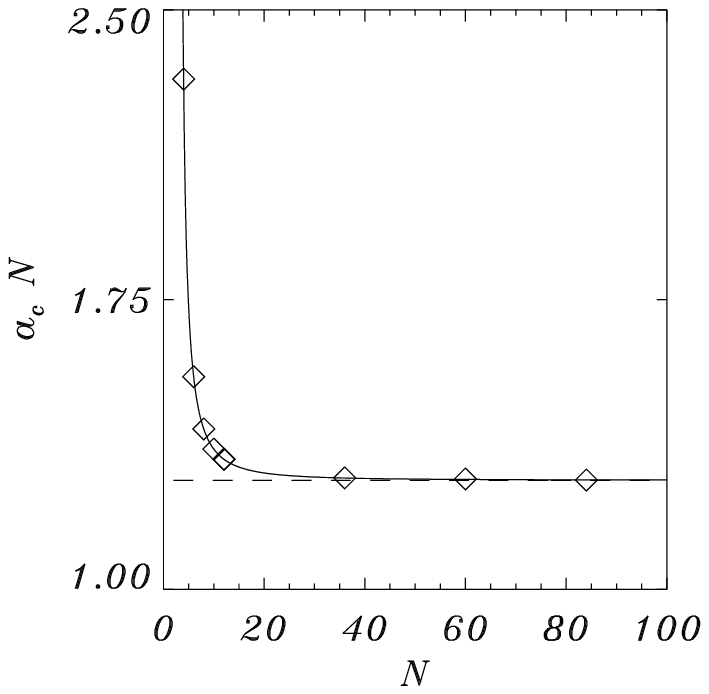,width=0.46\linewidth}
\epsfig{file=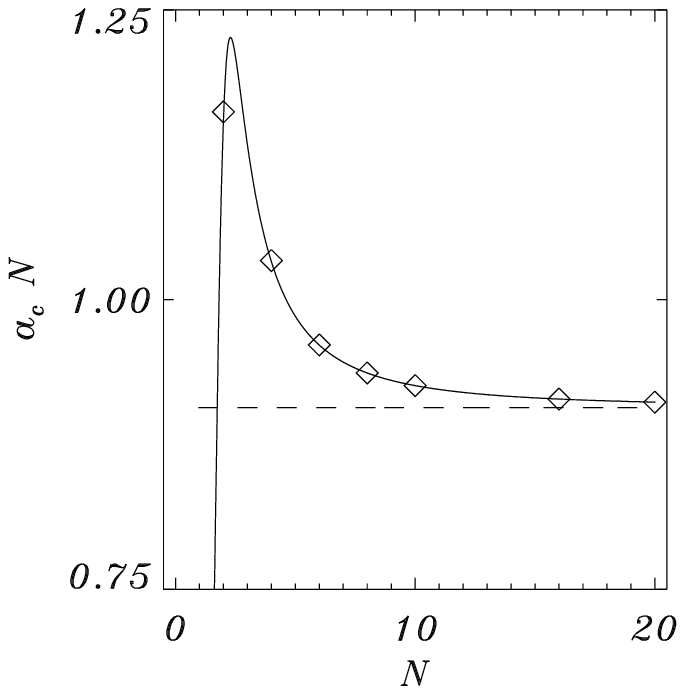,width=0.45\linewidth}} \caption{Left panel:
Modulational instability threshold amplitude for the $\pi$-mode
versus the number of particles in the one-dimensional FPU lattice.
The solid line corresponds to the analytical
formula~(\ref{final1d}), the dashed line to its large
$N$-estimate~(\ref{final1dapp}) and the diamonds are obtained from
numerical simulations. Right panel: Modulation instability threshold
for the $(\pi,\pi)$ mode versus number of oscillators in two
dimensional array. Solid line is derived from the exact analytical
consideration (\ref{final}), dashed line describes the estimate from
NLS equation in large $N$ limit (\ref{form}) and diamonds are result
of numerical simulations.} \label{modinstab1d}
\end{figure}

A first interesting case is $q=\pi$. One can easily see that
Eq.~(\ref{relatdispercorr}) admits two real and two complex conjugate
imaginary solutions if
and only if
\begin{equation}
\cos^2{Q\over 2}>{1+\alpha\over 1+3\alpha}.
\label{acrit}
\end{equation}
This formula was first obtained by Sandusky and Page (Eq.~(22) in
Ref.~\cite{sandusky}) using the rotating wave approximation.  The
first mode to become unstable when increasing the amplitude $a$
corresponds to the wavenumber $Q={2\pi/N}$.  Therefore, the critical
amplitude $a_c$ above which the $q=\pi$-mode looses stability is
\begin{equation}
a_c= \left( {\sin^2\left({\pi/ N}\right)\over 3
\left[3\cos^2\left({\pi/ N}\right)-1\right]} \right)^{1/2}\label{final1d}.
\end{equation}
This formula is valid for all even values of $N$ and its large
$N$-limit is
\begin{equation}
a_c= \frac{\pi}{\sqrt{6}N}+O\left({1\over {N^3}}\right)\label{final1dapp}.
\end{equation}
In Fig.~\ref{modinstab1d}, we show its extremely good agreement
with the critical amplitude determined from numerical simulations.
It is interesting to emphasize that the analytical
formula~(\ref{final1d}) diverges for $N=2$, predicting that the
$\pi$-mode is stable for all amplitudes in this smallest lattice.
This is in agreement with the Mathieu equation analysis  (see
Ref.~\cite{PR} p.~265).

It is also interesting to express this result in terms of the total
energy to compare with what has been obtained using other methods
\cite{zakharov,bountis,berman,flach,PR}. Since for the $\pi$-mode the
energy is given by $E=N(2a^2+4a^4)$, we obtain the critical
energy
\begin{equation}
E_c={2N\over 9}\sin^2\left({\pi\over
N}\right){7\cos^2\left({\pi/ N}\right)-1
\over \left[3\cos^2\left({\pi/ N}\right)-1\right]^2}.
\label{eqnc}
\end{equation}
For large $N$, we get
\begin{equation}
E_c={\pi^2\over 3N}+O\left({1\over {N^3}}\right).\label{scalingec}
\end{equation}
This asymptotic behavior is the same as the one obtained using the
narrow packet approximation in the context of the nonlinear
Schr{\"o}dinger equation by Berman and Kolovskii (Eq.~(4.1) in
Ref.~\cite{berman}). The correct scaling behavior with $N$ of the
critical energy has been also obtained by Bundinsky and Bountis
(Eq.~(2.22) in Ref.~\cite{bountis}) by a direct linear stability
analysis of the $\pi$-mode. The correct formula, using  this latter method,
has been independently obtained by Flach (Eq.~(3.20)) in
Ref.~\cite{flach}) and Poggi and Ruffo (p. 267 of Ref.~\cite{PR}).
Recently, the $N^{-1}$-scaling of formula~(\ref{scalingec}) has been confirmed
using a different numerical method and, interestingly, it holds also for
the  $2\pi/3$ and $\pi/2$ modes~\cite{Cafarella}.

This critical energy is also very close to the Chirikov
``stochasticity threshold'' energy obtained by the resonance overlap
criterion for the zone boundary mode\cite{izraeilchirikov}. The
stochasticity threshold phenomenon has been thoroughly studied for
long wavelength initial conditions, and it has been clarified that it
corresponds to a change in the scaling law of the largest Lyapunov
exponent\cite{pettini}.  We will show in Section \ref{localization1d}
that above the modulational instability critical energy for the
$\pi$-mode one reaches asymptotically a chaotic state with a positive Lyapunov
exponent, consistently with Chirikov's result.

The above results can be generalized to nonlinearities of $2\ell+1$
order in the equations of motion~(\ref{sub}). We limit the analysis
to the $\pi$-mode, for which the instability condition~(\ref{acrit})
takes the form
\begin{equation}
\cos^2{Q\over 2}>{1+\alpha\over 1+(2\ell+1)\alpha}, \label{acritnew}
\end{equation}
where
\begin{equation}
\alpha=\frac{(2\ell+1)!}{\ell ! (\ell+1) !}\, a^{2\ell}.
\end{equation}
Hence the critical amplitude above which the
$\pi$-mode is unstable is
\begin{equation}
a_c= \left[ {\ell ! (\ell+1) !\, \sin^2\left({\pi/ N}\right)\over
(2\ell +1) ! \left[(2\ell +1)\cos^2\left({\pi/ N}\right)-1\right]}
\right]^{1/2}\label{final1dpourp},
\end{equation}
leading to the large $N$ scaling
\begin{eqnarray}
a_c&\sim& N^{-{1/ \ell}}\\
E_c&\sim& N^{1-2/\ell}.
\end{eqnarray}
This scaling also corresponds to the one found in Ref.~\cite{Kladko}
when discussing tangent bifurcations of band edge plane waves in
relation with energy thresholds for discrete breathers. Their
``detuning exponent'' $z$ has a direct connection with the
nonlinearity exponent $\ell=z/2$. We will see in
Section~\ref{modinst2D} that this analogy extends also to higher
dimensions.

For fixed $N$, $a_c$ is an increasing function of the power of the
coupling potential with the asymptotic limit $\lim_{\ell\to\infty}
a_c=0.5$. Therefore, in the hard potential limit the critical energy
for the $\pi$-mode increases proportionally to $N$. The fact that we
find a higher energy region where the system is chaotic is not in
contradiction with the integrability of the one-dimensional system
of hard rods\cite{refhardrod}, because in the present case we have
also a harmonic contribution at small distances.

For the FPU-$\alpha$ model (quadratic nonlinearity in the equations of
motion), the $\pi$-mode is also an exact solution which becomes unstable
at some critical amplitude which, contrary to the case of the FPU-$\beta$
model, is $N$-independent\cite{sandusky,chechin}; which means that the
critical energy is proportional to $N$ and then that $\pi$-mode can be
stable in some low energy density limit also in the thermodynamic
limit.

It has also been
realized~\cite{PR,marie,julien,chechin,Shinohara,Rink2003} that group
of modes form sets which are invariant under the dynamics.  The
stability analysis~\cite{julien,chechin2004} of pair of modes has
shown a complex dependence on their relative amplitudes.  The
existence of such invariant manifolds has also allowed to construct
Birkhoff-Gustavson normal forms for the FPU model, paving the way to
KAM theory~\cite{Rink2000}.

\subsection{Higher dimensions}
\label{modinst2D}

In this Section, we will first discuss modulational instability of the
two-dimensional FPU model. The method presented in
Section~\ref{modinst1D} can be easily extended and the global physical
scenario is preserved. However, the scaling with $N$ of the critical
amplitude changes in such a way to make  critical energy constant, in
agreement with the analysis of Ref.~\cite{Kladko}.

The masses lie on a two-dimensional square lattice with unitary
spacing in the $(x,y)$ plane. We consider a small relative
displacement $u_{n,m}$ ($n,m\in [1,N]$) in the vertical direction
$z$.  Already with an harmonic potential, if the spring length at
equilibrium is not unitary, the series expansion in $u_{n,m}$ of the
potential contains all even powers. We retain only the first two terms
of this series expansion. After an appropriate rescaling of time and
displacements to eliminate mass and spring constant values, one gets
the following adimensional equations of motions
\begin{eqnarray}
\ddot u_{n,m}&=&u_{n+1,m}+u_{n-1,m}+u_{n,m+1}+u_{n,m-1}-4u_{n,m} \nonumber \\ &&
+\left(u_{n+1,m}- u_{n,m}\right)^3+ \left(u_{n-1,m}-u_{n,m}\right)^3
+\left(u_{n,m+1}- u_{n,m}\right)^3+\left(u_{n,m-1}- u_{n,m}\right)^3 .\label{2Dfpu}
\end{eqnarray}
Considering periodic boundary conditions, plane waves solutions have
the form
\begin{equation}
u_{n,m}=a\cos\left(q_xn+q_ym-\omega t\right). \label{harmonic}
\end{equation}
In the rotating wave approximation\cite{sandusky}, one immediately
obtains the dispersion relation
\begin{eqnarray}
\omega^2=4\sin^2\frac{q_x}{2}+4\sin^2\frac{q_y}{2}+12a^2\left[\sin^4
\frac{q_x}{2}+\sin^4\frac{q_y}{2}\right],
\label{dispersion}
\end{eqnarray}
which becomes exact for the zone-boundary mode $(q_x, q_y)=(\pi,\pi)$,
\begin{equation}
\omega_{\pi,\pi}^2=8(1+3a^2).
\label{2ddisp}
\end{equation}
To study the stability of the zone-boundary mode, we adopt a slightly
different approach. Namely, we consider the perturbed
relative displacement field of the form
\begin{equation}
u_{n,m}=\left(\frac{a}{2}+b_{n,m}\right)e^{i(\pi n+\pi m-\omega_{\pi,\pi}t)}+
c.c.,
\label{disp}
\end{equation}
where $b_{n,m}$ is complex. This approach turns out to be equivalent
to the one of Section~\ref{modinst1D} in the linear limit.

Substituting this perturbed displacement field in Eqs.~(\ref{2Dfpu}), we obtain
\begin{eqnarray}
[1+2\alpha]\left[ b_{n+1,m}+ b_{n-1,m}+ b_{n,m+1}+ b_{n,m-1}+ 4b_{n,m}\right] &&
\nonumber \\
-\alpha\left[b^*_{n+1,m}+ b^*_{n-1,m}+ b^*_{n,m+1}+ b^*_{n,m-1}+
4b^*_{n,m}\right]&&
=-\ddot b_{n,m}+2i\omega_{\pi,\pi}\dot b_{n,m}+\omega_{\pi,\pi}^2b_{n,m},
\label{bnm}
\end{eqnarray}
where $\alpha=3a^2$. Looking for  solutions of the form
\begin{equation}
b_{n,m}=Ae^{i(Q_xn+Q_ym-\Omega t)}+ Be^{-i(Q_xn+Q_ym-\Omega t)},
\end{equation}
we arrive at the following set of linear algebraic equations for the complex
constants $A$ and $B$
\begin{eqnarray}
\left[(\Omega+\omega_{\pi,\pi})^2-
8(1+2\alpha)\Delta\right]A+8\alpha \Delta B&=&0  \\
8\alpha \Delta A+ \left[(\Omega-\omega_{\pi,\pi})^2-
8(1+2\alpha)\Delta\right]B&=&0,
\label{eq}
\end{eqnarray}
where $2\Delta=\cos^2(Q_x/2)+\cos^2(Q_y/2)$. As for the
one-dimensional case, we require that the determinant of this linear
system in $A$ and $B$ vanishes, which leads to the following condition
\begin{eqnarray}
\left[(\Omega+\omega_{\pi,\pi})^2-8\Delta (1+2\alpha)\right]
\left[(\Omega-\omega_{\pi,\pi})^2-
8\Delta(1+2\alpha)\right] = 64\alpha^2\Delta^2.
\label{relatdispercorr2D}
\end{eqnarray}
This equation admits two real and two complex conjugated
imaginary solutions in $\Omega$ if
\begin{equation}
\Delta>\frac{1+\alpha }{1+3\alpha },
\end{equation}
which is the analogous of condition (\ref{acrit}) for two dimensions.
One can achieve the minimal nonzero value of the r.h.s. of the above expression choosing
$Q_x=0$, $Q_y=2\pi/{N}$, which leads to the following result for the
critical amplitude
\begin{equation}
a_c=\left(\frac{\sin^2(\pi/{N})}{3[3\cos^2(\pi/{N})+1]}\right)^{1/2}.
\label{final}
\end{equation}
Its large $N$ limit is
\begin{equation}
a_c= \frac{\pi}{\sqrt{12}N}+O\left({1\over {N^3}}\right)\label{final2dapp}.
\end{equation}
This prediction is compared with numerical data in
Fig.~\ref{modinstab1d}.
The agreement is good for all values of $N$.

Since the relation between energy and amplitude is now
$E=2N^2(2a^2+4a^4)$,
we obtain the critical energy in the large $N$-limit as
\begin{equation}
E_c={\pi^2\over 3}+O\left({1\over {N^2}}\right).\label{scalingec2D}
\end{equation}
This shows that the critical energy is now constant in the
thermodynamic limit, which agrees with the remark of Ref.~\cite{Kladko}
about the existence of a minimal energy for breathers formation~\cite{reviewbreather}.

The results of this Section can be easily extended to any dimension
$d$.  Repeating the same argument, we arrive at the following estimates
for the critical amplitude and energy in the large $N$ limit
\begin{eqnarray}
a_c&=& \frac{\pi}{\sqrt{6d}}\frac{1}{N}+O\left({1\over {N^3}}\right)\label{acinfini}\\
E_c&=&{\pi^2\over 3}N^{d-2}+O\left({N^{d-4}}\right).\label{ecinfini}
\end{eqnarray}
This means that the critical energy density $ \varepsilon_c=E_c/N$ for destabilizing the
zone boundary mode vanishes as $1/N^2$, independently of dimension.


\subsection{Large $N$ limit using the Nonlinear Schr{\"o}dinger equation}

The large $N$ limit expressions~(\ref{acinfini}) and~(\ref{ecinfini})
can be derived also by continuum limit considerations. We will derive
the general expression for any dimension $d$. The
displacement field can be factorized into a complex envelope part $\psi$
multiplied by the zone boundary mode pattern in $d$ dimensions.
\begin{equation}
u_{n_1,\dots,n_d }=\frac{\psi(n_1,\dots,n_d,t)}{2}e^{i\left(\pi \sum_{ i=1}^dn_i-\omega_{\pi,
    \dots , \pi }t\right)}+ c.c.,
\label{dispp}
\end{equation}
where
\begin{equation}
\omega_{\pi,\dots , \pi }=\sqrt{d\left[4(1+3|\psi|^2)\right]}.
\label{dispo}
\end{equation}
Substituting Eq.~(\ref{dispp}) into the FPU lattice equations in $d$ dimensions,
a standard procedure~\cite{RemoissenetSemiDis,book} leads to the following
$d$ dimensional Nonlinear Schr{\"o}dinger (NLS) equation:
\begin{equation}
i \frac{\partial \psi }{\partial t} + \frac{P}{2}\,
\Delta_d \psi- Q \psi
{\left| \psi \right|}^2 =0,
\label{nls}
\end{equation}
where $\Delta_d$ is the $d$ dimensional Laplacian. The parameters $P$ and
$Q$ are derived from the nonlinear dispersion relation
\begin{eqnarray}
\omega^2=\sum_{i=1}^{d}\left[4\sin^2\frac{q_{i}}{2}+12|\psi|^2\sin^4
\frac{q_{i}}{2}\right],
\label{dispersionddim}
\end{eqnarray}
as
\begin{eqnarray}
P&=&\frac{\partial^2\omega}{\partial
q_i^2}\left({q_1=\pi,\dots, q_d=\pi,|\psi| =0}\right)=\frac{1}{2\sqrt{d}}  \\
Q&=&-\frac{\partial\omega}{\partial
|\psi|^2}\left({q_1=\pi,\dots, q_d=\pi,|\psi| =0}\right)=-3\sqrt{d}.
\label{para}
\end{eqnarray}

Assuming that, at the first stage, modulation instability develops
along a single direction $x$ and that the field remains constant along all
other directions, one gets  the one-dimensional NLS equation
\begin{equation}
i \frac{\partial \psi }{\partial t} +\frac{P}{2}
\frac{\partial^2 \psi }{\partial x^2}- Q \psi
{\left| \psi  \right|}^2 =0.
\label{nls1}
\end{equation}
Following the results of the inverse scattering approach
\cite{zakharov}, any initial distribution of amplitude $|\psi|$ and
length $\lambda $ along $x$, and constant along all other directions,
produces a final localized distribution if \cite{lukomskii}
\begin{equation}
(|\psi|\lambda)^2>\pi^2\left|\frac{P}{Q}\right|.
\label{form0}
\end{equation}
This means that if the initial state is taken with constant amplitude
$|\psi|=a$ on the $d$-dimensional lattice with $N^d$ oscillators, the
modulation instability threshold is
\begin{equation}
(a_c{N})^2=\frac{\pi^2}{6d}
\label{form}
\end{equation}
which coincides with the leading order inEq. (\ref{acinfini}).

\section{Emergence of Localizations in Anti-FPU}
\label{localization1d}
\subsection{Conservative Case: Chaotic Breathers}
\label{chaotic}

In this Section, we will discuss what happens when the
modulational energy threshold is overcome. The first thorough
study of this problem can be found in Ref.~\cite{maledetirusi},
many years after the early pioneering work of Zabusky and
Deem~\cite{ZabuskyDeem}. Already in Ref.~\cite{maledetirusi}, it
has been remarked that an energy localization process takes place,
which leads to the formation of breathers~\cite{reviewbreather}.
This process has been further characterized in terms of
time-scales to reach energy equipartition and quantitative
localization properties in Ref.~\cite{cretegny}. The localized
structure which emerges after modulational instability has been
here called ``chaotic breather'' (CB).  The connection between CB
formation and continuum equations has been discussed in
Refs.~\cite{kosevichlepri,mirnov}, while the relation with the
process of relaxation to energy equipartition has been further
studied in Ref.~\cite{ulman}.  We will briefly recall some
features of the localization process in one dimension and present
new results for two dimensions.
\begin{figure}
\centerline{\epsfig{file=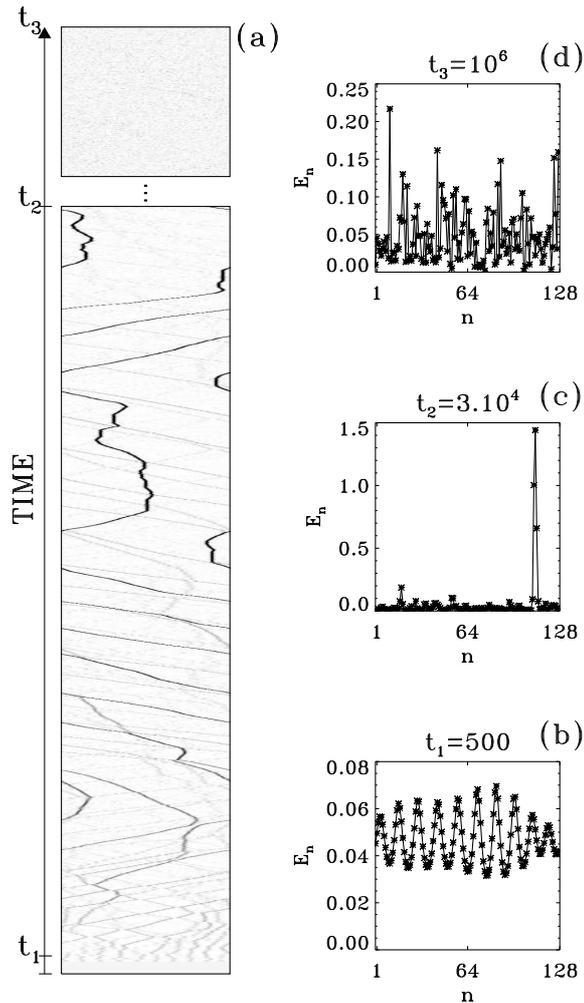,width=0.5\linewidth}} \caption{Time
evolution of the local energy (\ref{localenergy}). In panel~(a), the
horizontal axis indicates lattice sites and the vertical axis is
time. The grey scale goes from $E_n=0$ (white) to the maximum
$E_n$-value (black). The lower rectangle corresponds to $0<t<3000$
and the upper one to $5.994\ 10^5<t<6.10^5$. Figs.~(b), (c) and~(d)
show the instantaneous local energy $E_n$ along the $N=128$ chain at
three different times. Remark the difference in vertical amplitude
in panel~(c), when the CB is present. The initial $\pi$-mode
amplitude is $a=0.126>a_c\simeq0.010$.} \label{greyscales}
\end{figure}

For long time simulations, we use appropriate symplectic
integration schemes in order to preserve as far as possible the
Hamiltonian structure. For the one dimensional FPU, we adopt a
6th-order Yoshida's algorithm~\cite{yoshida} with a time step $dt
= 0.01$; this choice allows us to obtain an energy conservation
with a relative accuracy $\Delta E / E$ ranging from $10^{-10}$ to
$10^{-12}$. For two dimensions, we use instead the 5--th order
symplectic Runge--Kutta--Nystr\"{o}m algorithm of
Ref.~\cite{Calvo}, which gives a similar quality of energy
conservation.

\begin{figure}
\centerline{\epsfig{file=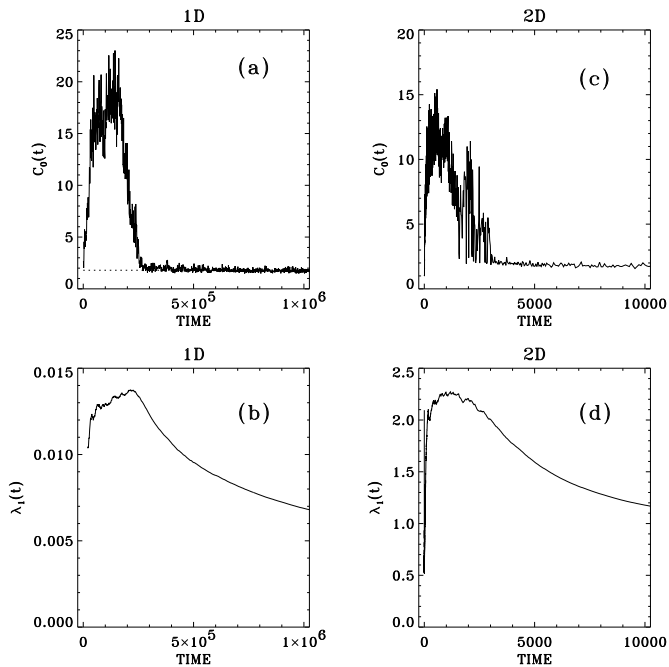,width=0.6\linewidth}}
\caption{Panel~(a) presents the evolution of $C_0(t)$ of formula
  (\ref{clocal}) for the one-dimensional FPU lattice with $N=128$
  oscillators, initialized on the $\pi$-mode with an amplitude
  $a=0.126>a_c\simeq0.010$. The dashed line indicates the equilibrium value
  $\bar C_0=1.795$.  Panel~(b) presents the corresponding finite time
  largest Lyapunov exponent. Panel~(c) shows $C_0(t)$ for the
  two-dimensional FPU lattice with $20*20$ oscillators, initialized on
  the $(\pi,\pi)$-mode with an amplitude $a=0.425>a_c\simeq0.045$.
  Panel~(d) presents the finite time largest Lyapunov exponent for two
  dimensions.}
\label{cohlyap}
\end{figure}

We report in Fig.~\ref{greyscales}(a) a generic evolution of the
one dimensional $\pi$-mode above the  modulation instability
critical amplitude ($ a > a_c$). The grey scale refers to the
energy residing on site~$n$,
\begin{equation}
E_n = {1\over 2} \dot{u}_n^2 + {1\over 2}V(u_{n+1}-u_n) +
{1\over 2} V(u_n-u_{n-1}),\label{localenergy}
\end{equation}
where the FPU-potential is $V(x) = {1\over 2}x^2 + {1\over 4} x^4$.
Figs.~\ref{greyscales}(b), \ref{greyscales}(c) and \ref{greyscales}(d)
are three successive snapshots of the local energy $E_n$ along the
chain. At short time, a slight modulation of the energy in the system
appears (see Fig.~\ref{greyscales}(b)) and the $\pi$-mode is
destabilized~\cite{sandusky}. Later on, as shown in
Fig.~\ref{greyscales}(a), only a few localized energy packets emerge:
they are breathers~\cite{reviewbreather}.  Inelastic collisions of breathers have a
systematic tendency to favour the growth of the big breathers at the
expense of small ones~\cite{prllocali,BangPeyrard}. Hence, in the
course of time, the breather number decreases and only one, of very
large amplitude, survives (see Fig.~\ref{greyscales}(c)): this is the
localized excitation we have called chaotic breather (CB). The CB
moves along the lattice with an almost ballistic motion: sometimes it
stops or reflects. During its motion the CB collects energy and its
amplitude increases. It is important to note that the CB is never at
rest and that it propagates with a given subsonic
speed~\cite{KosevichCorso}.  Finally,  the CB decays and the
system reaches  energy equipartition, as illustrated in
Fig.~\ref{greyscales}(d).

In order to obtain a quantitative characterization of
energy localization, we introduce the ``participation ratio''
\begin{equation}
C_0(t)=N {\displaystyle \sum_{i=1}^N E_i^2\over
\left(\displaystyle \sum_{i=1}^N E_i  \right)^2 },
\label{clocal}
\end{equation}
which is of order one if $E_i = E/N$ at each site of the chain and of
order $N$ if the energy is localized on only one site. In
Fig.~\ref{cohlyap}(a), $C_0$ is reported as a function of time.
Initially, $C_0$ grows, indicating that the energy, evenly distributed
on the lattice at $t=0$, localizes over a few sites.  This localized
state survives for some time. At later times, $C_0$ starts to decrease
and finally reaches an asymptotic value $ \bar C_0$ which is
associated with the disappearance of the CB (an estimate of $ \bar
C_0$ has been derived in Ref.~\cite{cretegny} taking into account
energy fluctuations and is reported with a dashed line in
Fig.~\ref{cohlyap}(a)).  At this stage, the energy distribution in
Fourier space is flat, i.e.  a state of energy equipartition is
reached.

In Fig.~\ref{cohlyap}(b), we show the finite time largest Lyapunov
exponent $\lambda_1(t)$ for the same orbit as in Fig.~\ref{cohlyap}(a). We
observe a growth of $\lambda_1(t)$ when the CB emerges on the lattice and a
decrease when it begins to dissolve. The peak in $\lambda_1(t)$ perfectly
coincides with the one in $C_0$. Due to this increase of chaos
associated with localization, we have called the breather chaotic
(although chaos increase could be the result of more complicated
processes of interaction with the background).

In Ref.~\cite{cretegny}, the time-scale for the relaxation to
equipartition has been found to increase as $(E/N)^{-2}$ in the
small energy limit. This has been confirmed by the followers of this
study~\cite{kosevichlepri,ulman,mirnov}. Such power law scalings are
found also for the FPU relaxation starting from long
wavelengths~\cite{delucca}: the so-called {\it FPU problem}.  We
have termed the relaxation process which starts from short
wavelengths the {\it Anti-FPU problem}, just because of the
similarities in the scaling laws. The main feature of the latter
problem is that relaxation to equipartition goes through a complex
process of localized structures formation well described by
breathers or, in the low-amplitude limit, by solitons of the
NonLinear Schrodinger equation. On the contrary, for the original
FPU problem, an initial long wavelength excitation breaks up into a
train of mKdV-solitons. The final relaxation to equipartition is
however due to an energy diffusion process which has similar
features for both the FPU and the anti-FPU problem~\cite{ulman}.

\begin{figure}[ht]
\centerline{\epsfig{file=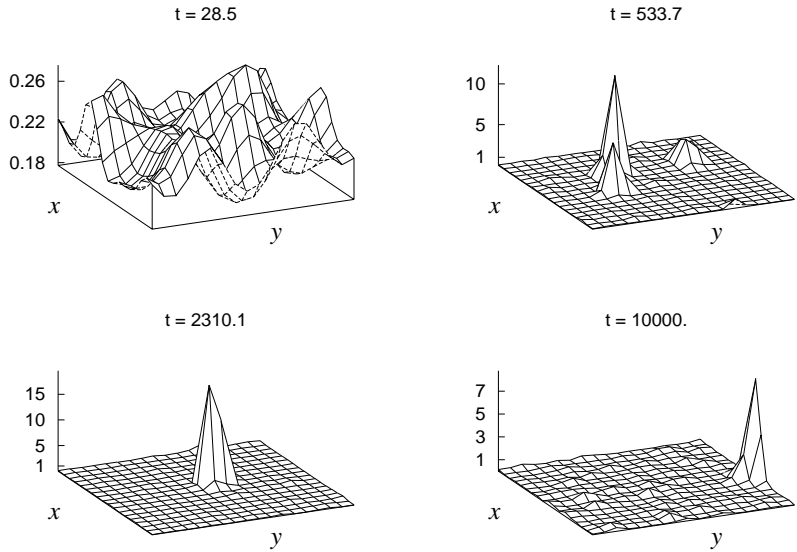,width=0.8\linewidth}}
\caption{Local energy (\ref{localenergy2d})  surface plots  for the
  two-dimensional FPU lattice with $20*20$ oscillators, initialized on
  the $(\pi,\pi)$-mode with an amplitude $a=0.225496>a_c\simeq0.0453450$.
  Snapshots at four different times $t$ are shown. Breathers form
  after a coalescence process similarly to the one-dimensional
  case. The mobility of the breathers is evident and one also observes
  in the last panel the final decrease.}
\label{fran1}
\end{figure}

A similar evolution of the local energy
\begin{eqnarray}
E_{n,m} = {1\over 2} \dot{u}_{n,m}^2 &&+ {1\over 4}V(u_{n+1,m}-u_{n,m})
+ {1\over 4}V(u_{n,m+1}-u_{n,m}) \nonumber\\
&&+ {1\over 4}V(u_{n-1,m}-u_{n,m}) + {1\over 4}V(u_{n,m-1}-u_{n,m}) \label{localenergy2d}
\end{eqnarray}
is observed for the two-dimensional case (see Fig. \ref{fran1}). In
this figure, we just show the initial evolution which leads to the
breathers formation. As for the one-dimensional case, bigger
breathers eat up smaller ones, and finally only two breathers
survive. We don't observe coalescence to a single breather because
collisions are more rare in two dimensions. After the formation of a
few localized structures, one also observes the final relaxation to
equipartition which is not shown in Fig. \ref{fran1}. This latter is
instead evident from the time evolution of $C_0(t)$, the
localization parameter, shown in Fig. \ref{cohlyap}(c): its behavior
is very similar to the one-dimensional case. Indeed, also the
largest finite time Lyapunov exponent behaves similarly (see Fig.
\ref{cohlyap}(d)).

\subsection{Nonconservative Case: Homogeneously Driven-Damped Anti-FPU}
\label{driven}

The equations of motion of the ``externally driven" and damped FPU
chain read as follows:
\begin{eqnarray}
\ddot u_n=u_{n+1}+u_{n-1}-2u_n +(u_{n+1}-u_n)^3+(u_{n-1}-u_n)^3 -\gamma
\dot u_n+f\cos (\omega t+p n),
\label{fpu}
\end{eqnarray}
where the forcing and damping strengths are gauged by the parameters
$f$ and $\gamma$, respectively; $\omega$~and $p$ are the driving
frequency and wavenumber. Considering the "Anti-FPU" situation, we
restrict ourselves to the case $\pi/2 <|p| <\pi$. Moreover, here we
present only the results concerning the range of driving frequencies
$|\omega|>\omega_p\equiv \sqrt{2(1-\cos p)}$ for which stationary
multibreather states develop (they are static if $p=\pi$ and move
for other cases). For other driving frequencies, one deals with
the stationary periodic patterns described in Refs.~\cite{ssr,ssrphysd}.

\begin{figure}
\centerline{\epsfig{file=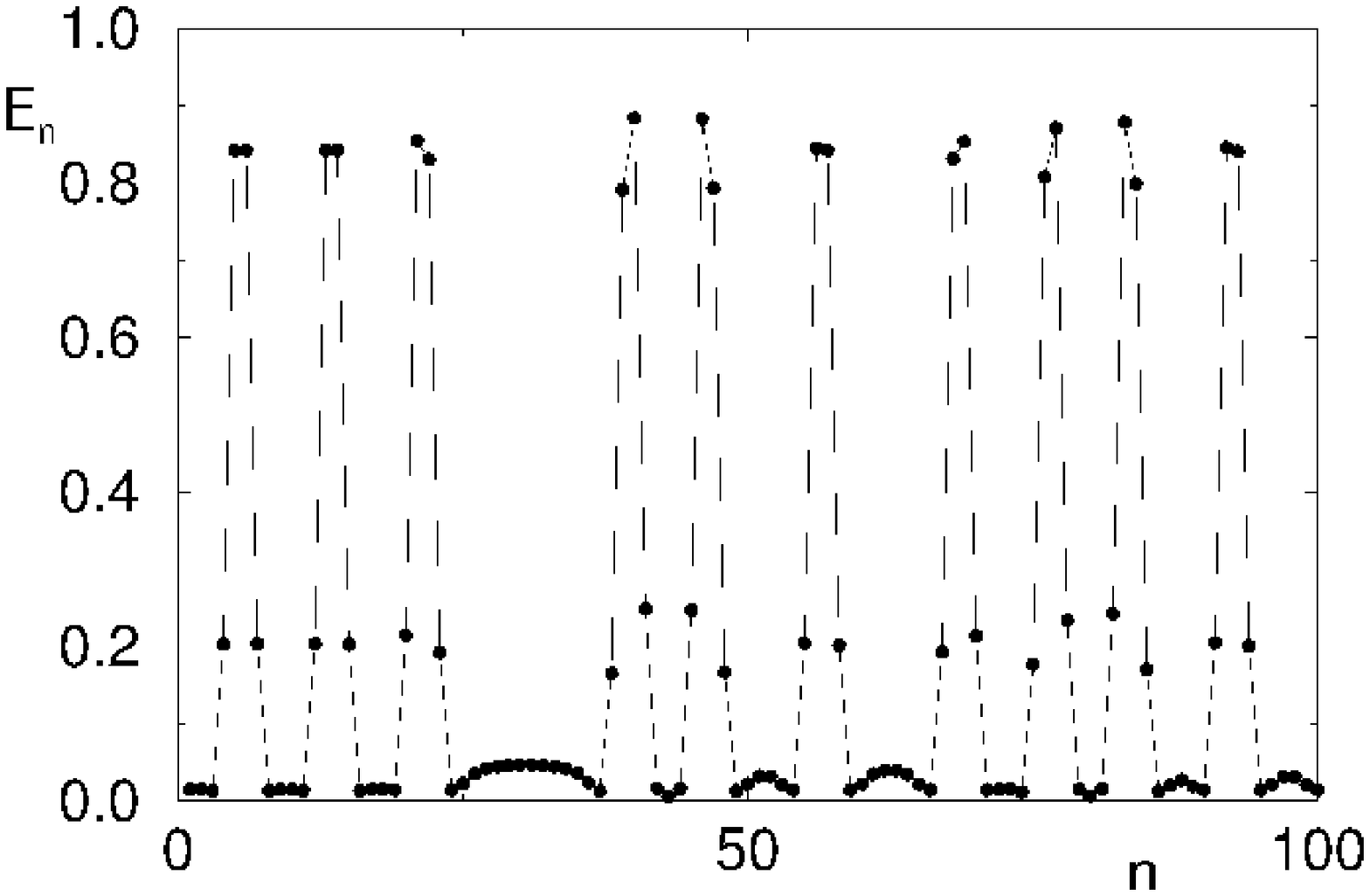,width=0.37\linewidth,angle=-90}\hskip.75truecm
\epsfig{file=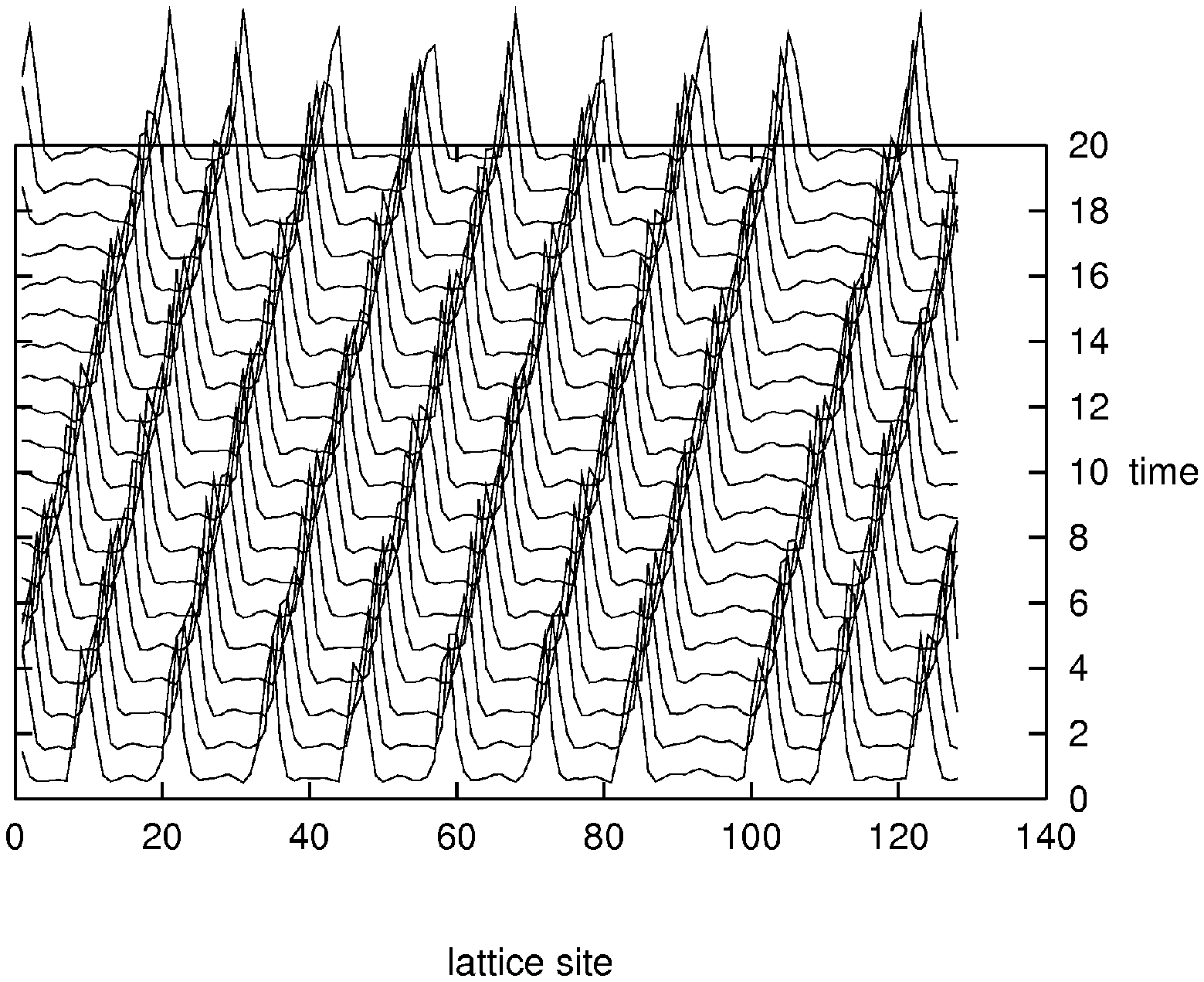,width=0.46\linewidth,angle=-90}} \caption{Left
panel: Static multibreather pattern generated after the modulational
instability for $p=\pi$ ($\omega=2.4$, $f=0.2250$). This pattern
stabilizes at $t \simeq 5\, 10^3$. Right panel: Travelling
multibreather pattern of the $p  \neq \pi$-mode for a lattice of
$N=512$ sites. Here, $\omega=2.05$, $f=0.075$ and $p=2.4542$.}
\label{multibre}
\end{figure}

\begin{figure}
\centering{\includegraphics[width=0.4\textwidth,angle=-90]{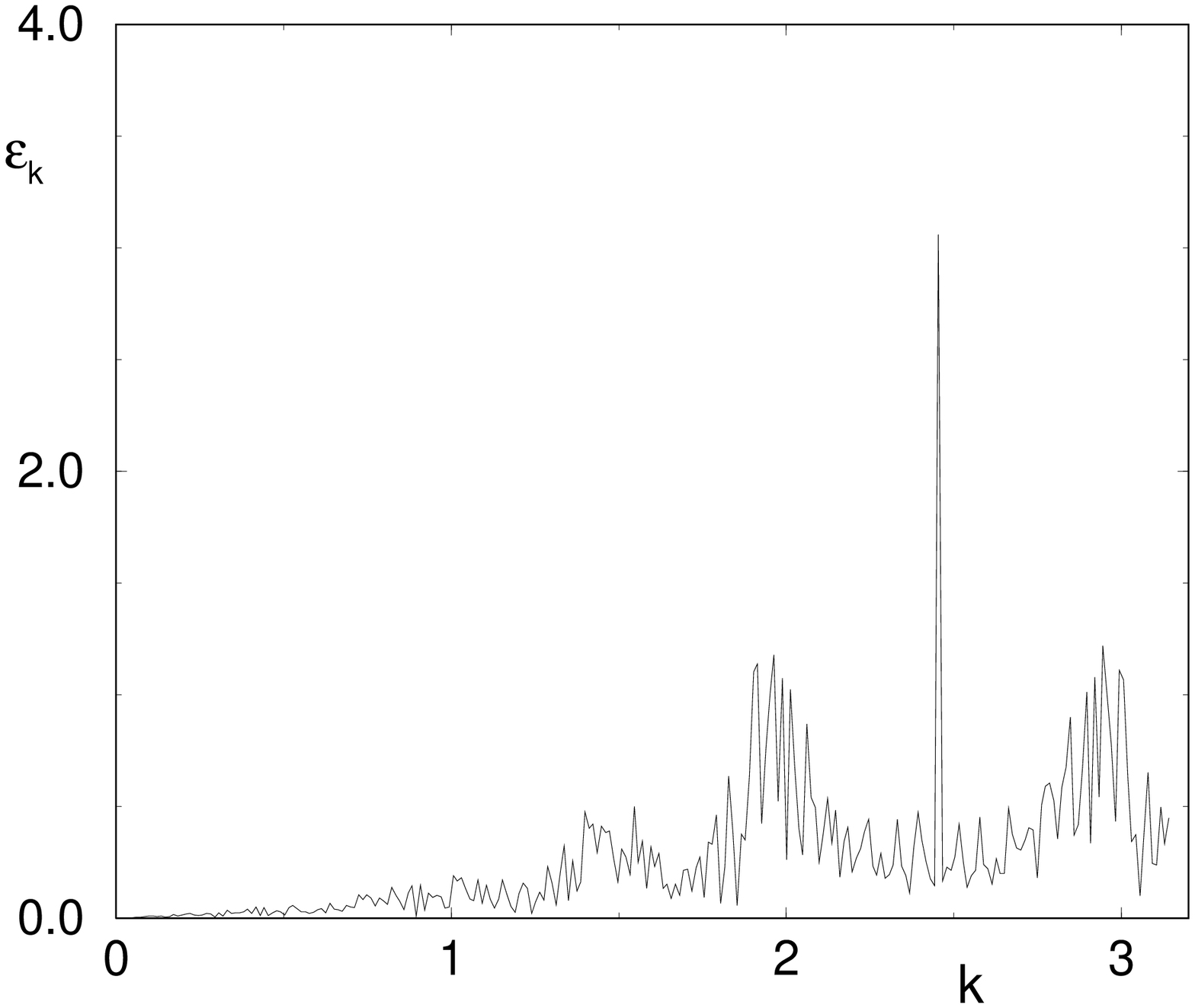}
\hspace{1cm}
\includegraphics[width=0.4\textwidth,angle=-90]{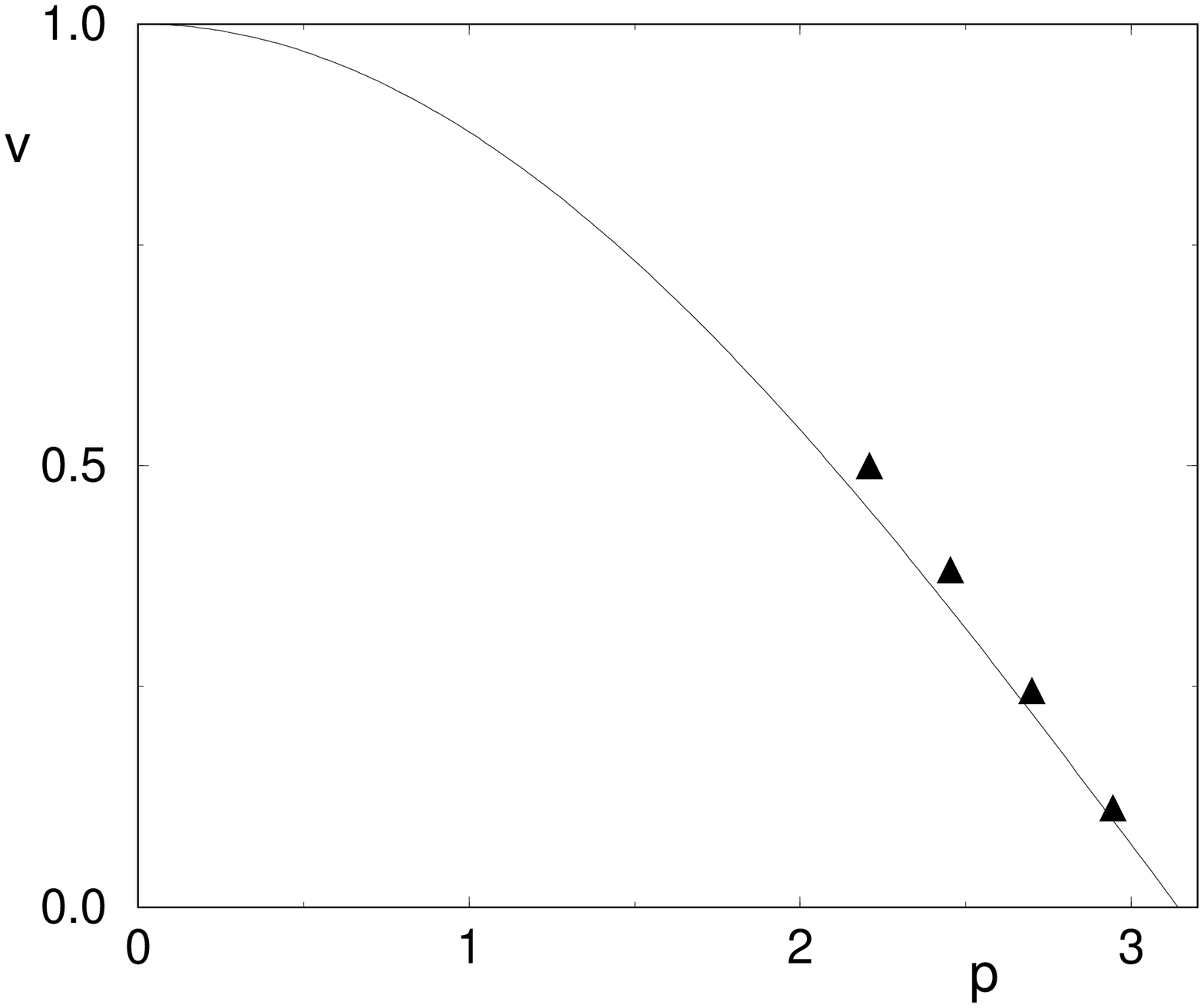}}
\caption{Left panel: Spatial spectrum of the travelling
multibreather. $\epsilon_k=|\dot{U}_k|+\omega^2|U_k|$, where $U_k$
is the $k$-th component of the Fourier spectrum of the displacement
field $u_n$. Same parameters as in Fig.~\ref{multibre}. Right panel:
Velocity (triangles) of travelling multibreathers  versus the
wavenumber of the forcing. The solid line is the group velocity of
linear waves.} \label{spectrum}
\end{figure}

Examples of appearance of either static ($p=\pi$) or traveling
($p\neq\pi$) multibreather states are shown in Fig.~\ref{multibre},
where we plot the local energy
vs. the lattice position sampled at the period of the forcing.
The corresponding spatial Fourier spectrum is shown in Fig.~\ref{spectrum}.
The broad band structure of the spectrum reflects the non perfect periodic
arrangement of the localized peaks in Fig.~\ref{multibre}.

Such states can be described in terms of soliton solutions of
an associated suitable driven-damped nonlinear Schr\"odinger (NLS) equation.
Let us first make the following definition:
\begin{equation}
u_n=\frac{1}{2}\left[a_p(n,t)e^{i(\omega t+pn)} +
a_{p}^+(n,t)e^{-i(\omega t+pn)}\right],
\label{aq1}
\end{equation}
where $a_p$ and its conjugate are smooth functions of $n$. Such an
assumption is possible if the wavepacket is concentrated around the
driving mode $\Delta k\ll\pi$. Substituting (\ref{aq1}) into the
equations of motion~(\ref{fpu}), one gets:
\begin{eqnarray}
(\omega_p^2-\omega^2)a_p && +2i\omega\left(\frac{\partial
a_p}{\partial t}- v\frac{\partial a_p}{\partial n}\right)
+\frac{\omega_p^2}{4}\frac{\partial^2 a_p}{\partial
n^2}+\frac{3}{4}\omega_p^4 a_p|a_p|^2=-i\omega\gamma a_p-f.
\label{ac}
\end{eqnarray}
After performing the following re-scalings
\begin{eqnarray}
t'=\frac{\omega^2-\omega_p^2}{2\omega}t, \qquad \xi=\frac{2\sqrt{\omega^2-\omega_p^2}}{\omega_p}
(n-vt),\nonumber \\
\Psi=\sqrt{\frac{3}{8}}\frac{\omega_p^2}{\sqrt{\omega^2-\omega_p^2}}e^{it'}a_p(\xi,t'),
\qquad \gamma'=\frac{\omega}{\omega^2-\omega_p^2}\gamma, \qquad
h=\sqrt{\frac{3}{8}}\frac{\omega_p^2}{(\omega^2-\omega_p^2)^{3/2}}f,\nonumber
\end{eqnarray}
and choosing a reference frame moving with velocity $v=\partial
\omega_p/\partial p=\sin p/\omega_p$, Eq.~(\ref{ac}) reduces to the
well studied ``externally" driven (or ac driven) damped NLS
equation~\cite{baras,baras2}:
\begin{equation}
i\frac{\partial\Psi}{\partial
t'}+\frac{\partial^2\Psi}{\partial\xi^2}+2\Psi|\Psi|^2
=-i\gamma'\Psi-he^{it'}. \label{nls2}
\end{equation}
Exact soliton solutions of this equation can be obtained for
$\gamma'=0$, see Eqs.~(37-40) of Ref.~\cite{baras}. Moreover,
multisoliton solutions are also derived in Ref.~\cite{baras2}. What
we observe in Fig.~\ref{multibre} might well be a superposition of
such solutions to form a train of ``intrinsically localized"
structures. However, one should bear in mind that NLS solutions can
describe only low amplitude states. Therefore, they can be only a
rough approximation of the pattern displayed in Fig.~\ref{multibre},
which shows high amplitude localized peaks.

In the right panel of Fig.~\ref{spectrum}, we plot the speed of the
travelling multibreather as a function of the wavenumber of the
forcing $p$-mode, which compares well with the group velocity of the
corresponding linear waves, showing that nonlinear effects are
negligible in this parameter range.

\subsection{Driving by one end: Ordinary and Bandgap Transmission}
\label{supra}

To simulate the effect of an impinging wave, we impose to the
$\beta$-FPU chain [$\ell=1$ in Eq.~(\ref{sub})], the boundary
condition
\begin{equation}
u_0(t)  =  A \cos \omega t, \label{drive}
\end{equation}
while free boundary conditions are enforced on the other side of the
chain.

In order to be able to observe a stationary state in the {\it
conducting} regime, we need to steadily remove the energy injected
in the lattice by the driving force.  Thus, we damp a certain number
of the rightmost sites (typically 10\% of the total) by adding a
viscous term $-\gamma \dot u_n$ to their equations of motion.  A
convenient indicator to look at is the averaged energy flux
$j=\sum_n j_n/N$, where the local flux $j_n$ is given by the
following formula~\cite{rep}
\begin{equation}
j_n=\frac12 (\dot u_n+\dot u_{n+1})\left[u_{n+1}-u_n+(u_{n+1}-u_{n})^3\right].
\label{flux}
\end{equation}
Time averages of this quantity are taken in order to characterize
the insulating (zero flux)/con\-duc\-ting (non zero flux) state of
the system.

\subsubsection{In-band driving: nonlinear phonons}

For illustration, we first discuss the case when the driving frequency is located
inside the phonon band. Although trivial, this issue is of importance to better
appreciate the fully nonlinear features described later on.

Under the effect of the driving (\ref{drive}), we can look for extended
quasi-harmonic solutions (nonlinear phonons) of the form
\begin{equation}
u_n     =A\cos(kn-\omega t). \label{0}
\end{equation}
We consider the semi--infinite chain, so that $k$ varies
continuously between $0$ and $2\pi$. The nonlinear dispersion
relation can be found in the rotating wave approximation (see e.g.
Ref.~\cite{rwa}). Neglecting higher--order harmonics, it reads
\begin{equation}
\omega_0^2(k,A) = 2(1-\cos k)+3(1-\cos k)^2 A^2. \label{2}
\end{equation}
Thus the nonlinear phonon frequencies range from 0 to the upper band--edge
$\omega_0(\pi,A)\geq 2$.

\begin{figure}[h]
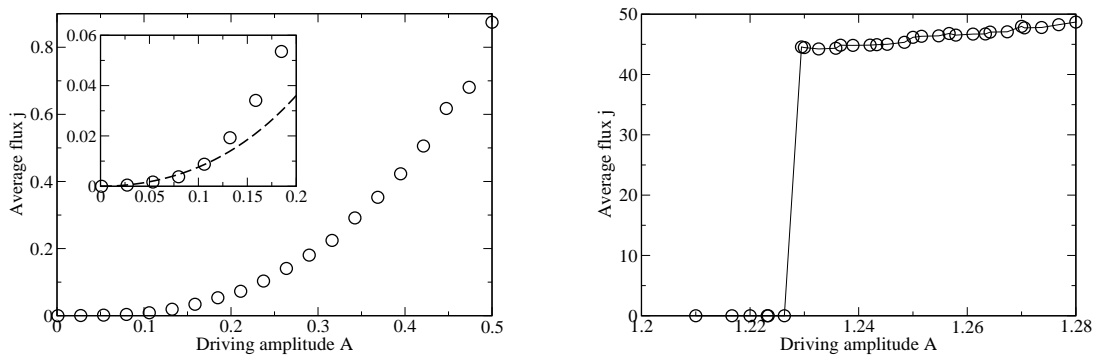

\begin{center}\leavevmode
\includegraphics[width=0.45\linewidth,clip]{Flux_A_W1.8.eps} \hspace{1cm}
\includegraphics[width=0.45\linewidth,clip]{Flux_A_W3.5.eps}
\end{center}
\caption{Left panel: Average energy flux vs.
driving amplitude for in-band forcing, $\omega=1.8$, $\gamma=5$.
Data have been averaged over $10^5$ periods of the driving. The
inset is an enlargment of the small-amplitude region and the dashed
line is the single nonlinear phonon approximation (\protect\ref{jphon}).
Right panel: Average energy flux vs.
driving amplitude for out--band forcing, $\omega=3.5$, $\gamma=5$.
Data have been averaged over $2\,10^5$ periods of
the driving for a chain of $N=512$ particles.}
\label{inband}
\end{figure}

If we simply assume that only the resonating phonons whose wavenumbers satisfy
the condition
\begin{equation}
\omega=\omega_0(k,A)
\end{equation}
are excited, we can easily estimate the energy flux.
Neglecting, for simplicity, the nonlinear force terms in the definition of the
flux (\ref{flux}), we have
\begin{equation}
j = \frac12 v(k,A) \, \omega^2 A^2, \label{jphon}
\end{equation}
where $v$ is the group velocity as derived from dispersion  relation
(\ref{2}).  This simple result is in very good agreement with
simulations, at least for small enough amplitudes (see left panel of
Fig.~\ref{inband}). For $A > 0.15$,  the measured flux is larger
than the estimate (\ref{jphon}), indicating that something  more
complicated occurs in the bulk (possibly, a multiphonon
transmission) and that higher-order nonlinear terms must be taken
into account.

\subsubsection{Out-band driving: supratransmission}

Let us now turn to the more interesting case in which the driving
frequency lies outside the phonon band, $\omega>\omega_0(\pi,0)=2$.
In a first series of numerical experiments, we have initialized the
chain at rest and switched on the driving at time $t=0$. To avoid
the formation of sudden shocks~\cite{shocks}, we have chosen to
increase smoothly the amplitude from 0 to the constant value $A$ at
a constant rate, i.e.
\begin{equation}
u_0=A\cos(\omega t)\left[1-e^{-t/\tau_1}\right], \label{dr1}
\end{equation}
where typically we set $\tau_1=10$.

At variance with the case of in--band forcing, we observe a sharp increase of the flux
at a given threshold amplitude of the driving, see right panel of Fig.~\ref{inband}.
This  phenomenon has been denoted as {\it nonlinear supratransmission}~\cite{leon} to
emphasize the role played by nonlinear localized excitations in triggering the energy flux.

This situation should be compared with the one of in--band driving, shown in
Fig.~\ref{inband}, where no threshold for conduction exists and the flux
increases continuously from zero (more or less quadratically
in the amplitude).
Indeed, the main conclusion that can be drawn from the previous section is that there
cannot be any amplitude threshold for energy transmission in the case of
in-band forcing. Moreover, although at the upper band edge the flux vanishes, since
it is proportional to the group velocity (see formula (\ref{jphon})), it is
straightforward to prove that it goes to zero with the square root of the distance to the band
edge frequency. Hence, the sudden jump we observe in the out-band case cannot
be explained by any sort of quasi-linear approximation.

In the following, we investigate the physical origin of nonlinear
supratransmission, distinguishing the cases of small and large
amplitudes.
\begin{figure}[b]
\begin{center}\leavevmode
\includegraphics[width=0.4\linewidth]{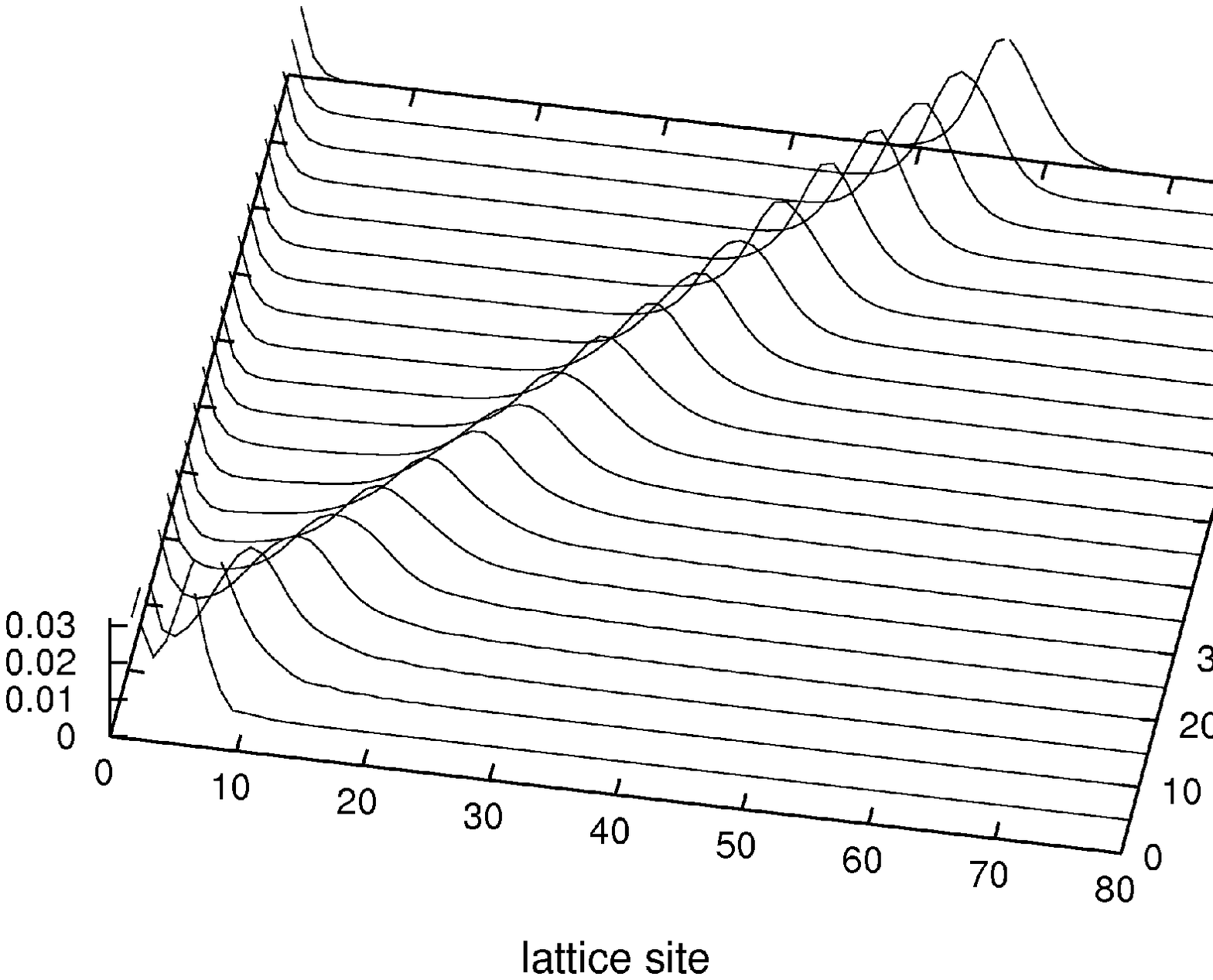} \hspace{1cm}
\includegraphics[width=0.5\linewidth]{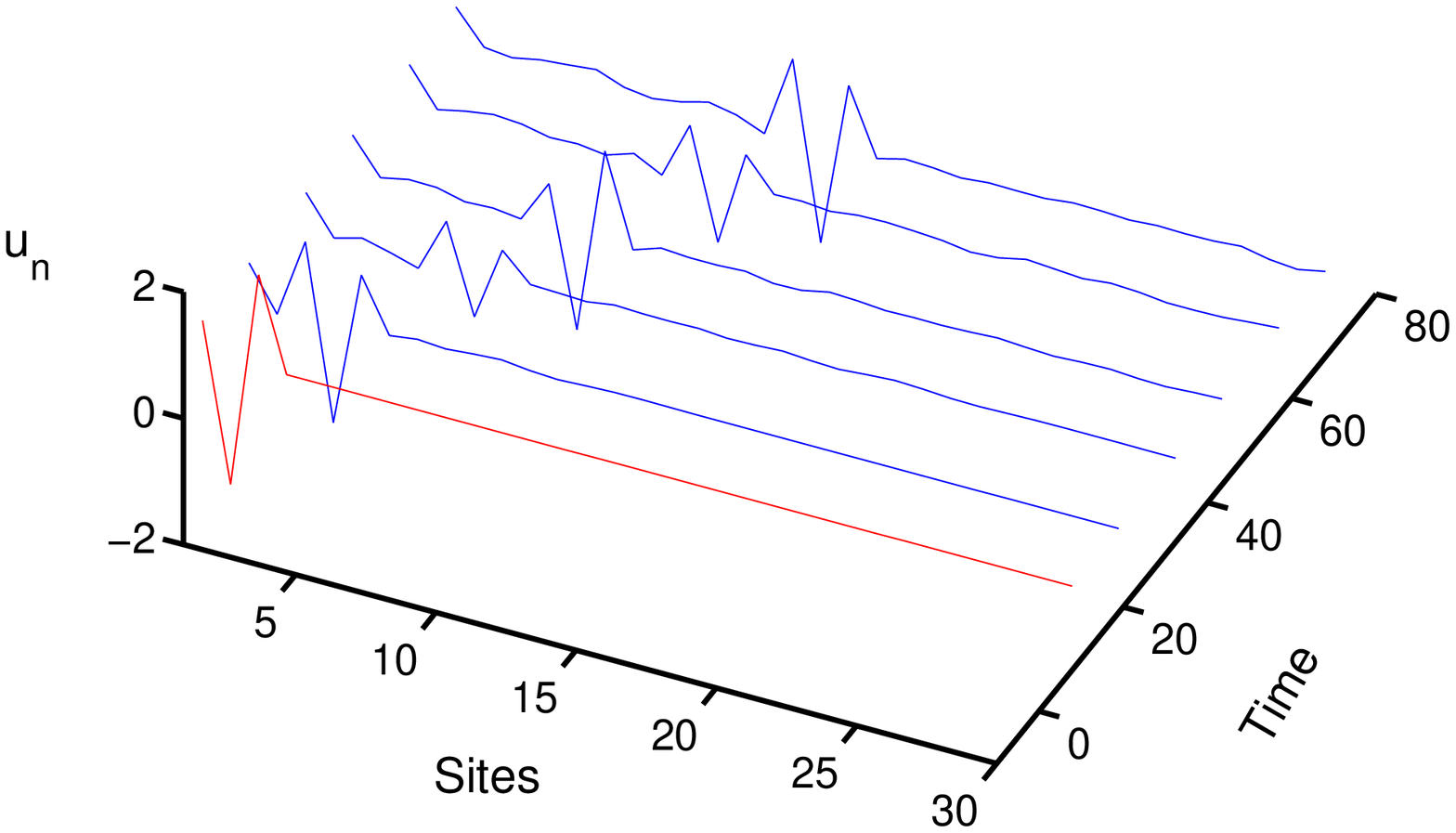}
\caption{Left panel: Snapshot of the local energy below the
supratransmission threshold $A=0.15 < A_{th}$ for $\omega =2.1$
$\gamma=10$. The initial condition is an envelope soliton
(\ref{sol}) with $x_0=+1.8$. Right panel: Snapshot of particle
displacements $u_n$ below the supratransmission threshold for a
driving frequency $\omega=5.12$ and a driving amplitude
$A=0.5<A_{th}=2.05$. One can observe, similarly to the left panel,
that a moving discrete breather appears at the left boundary and
propagates inside the bulk, leaving behind the static solution.}
\label{2brea}
\end{center}
\end{figure}

When the driving frequency is only slightly above the band ($0<\omega-2 \ll 1$),
one can resort to the continuum envelope approximation. Since we expect the
zone--boundary mode $k=\pi$ to play a major role, we let
\begin{equation}
u_n  =   (-1)^n \frac12 \left[ \psi_n\, e^{i\omega t} + \psi_n^*\,
e^{-i\omega t}\right].
\end{equation}
In the rotating wave approximation~\cite{rwa} and for slowly varying
$\psi_n$, one obtains from the FPU lattice equations the nonlinear
Schr\"odinger equation ($\psi_n \longrightarrow
\psi(x,t)$)~\cite{scott}
\begin{equation}
2i\omega \dot\psi  =  (\omega^2 - 4) \psi -\psi_{xx} -12
\psi|\psi|^2, \label{nlsnls}
\end{equation}
with the boundary condition $\psi(0,t)=A$.

The well-known {\it static} single--soliton solution of Eq.~(\ref{nlsnls}) corresponds to the
family of envelope solitons (low-amplitude discrete breathers)
\begin{equation}
u_n  =  a (-1)^n \cos(\omega t) \, {{\rm sech}\left[\sqrt{6}
(n-x_0)a\right]}, \label{sol}
\end{equation}
with amplitude $a=\sqrt{(\omega^2-4)/6}$. The maximum of the soliton shape is
fixed by the boundary condition to be
\begin{equation}
x_0 = \pm {{\rm acosh}(a/A) \over a\sqrt{6}}. \label{x0}
\end{equation}
In this approximation, we have two possible solutions: one with the
maximum outside the chain, which is purely decaying inside the chain
(minus sign in (\ref{x0})), and another with the maximum located
within the chain (plus sign in (\ref{x0})). Overcoming the
supratrasmission threshold corresponds to the disappearence of both
solutions. Indeed, when the driving amplitude reaches the critical
value $A_{th}$, given by
\begin{equation}
\omega^2=4+6A_{th}^2, \label{5}
\end{equation}
solution (\ref{sol}) ceases to exist.

We have investigated this issue by simulating the lattice dynamics
with the initial conditions given by Eqs.~(\ref{sol}) and
(\ref{x0}). The evolution of the local energy $E_n$ (see
Eq.~(\ref{localenergy}))
is shown in left panel of Fig.~\ref{2brea}. The solution with the maximum inside the chain
slowly moves towards the right and, eventually, leaves
a localized boundary soliton (with maximum outside the chain) behind. The release of energy to the
chain is non stationary and does not lead to a conducting state.

\begin{figure}[b]
\begin{center}\leavevmode
\includegraphics[width=0.4\linewidth]{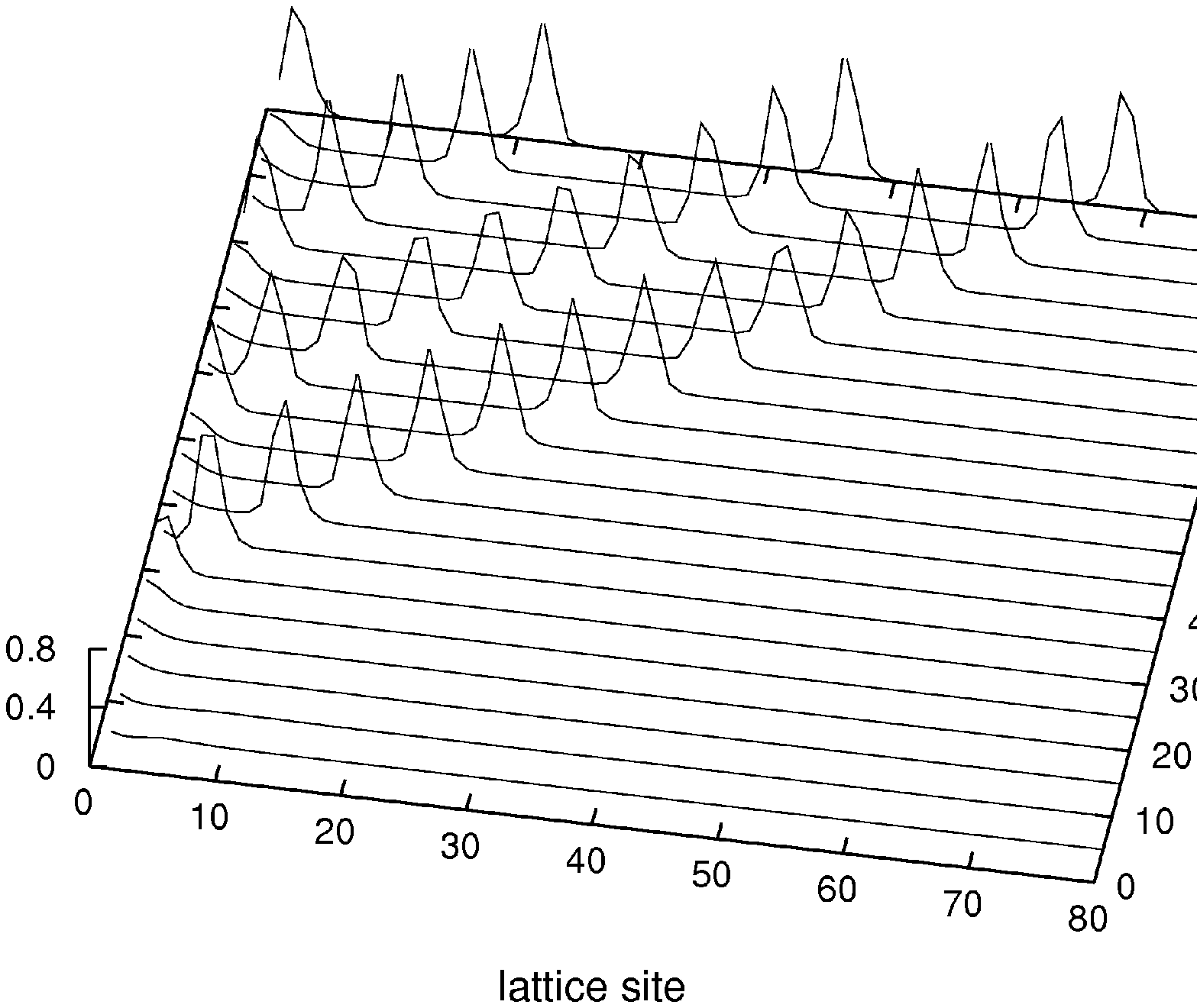} \hspace{1cm}
\includegraphics[width=0.5\linewidth]{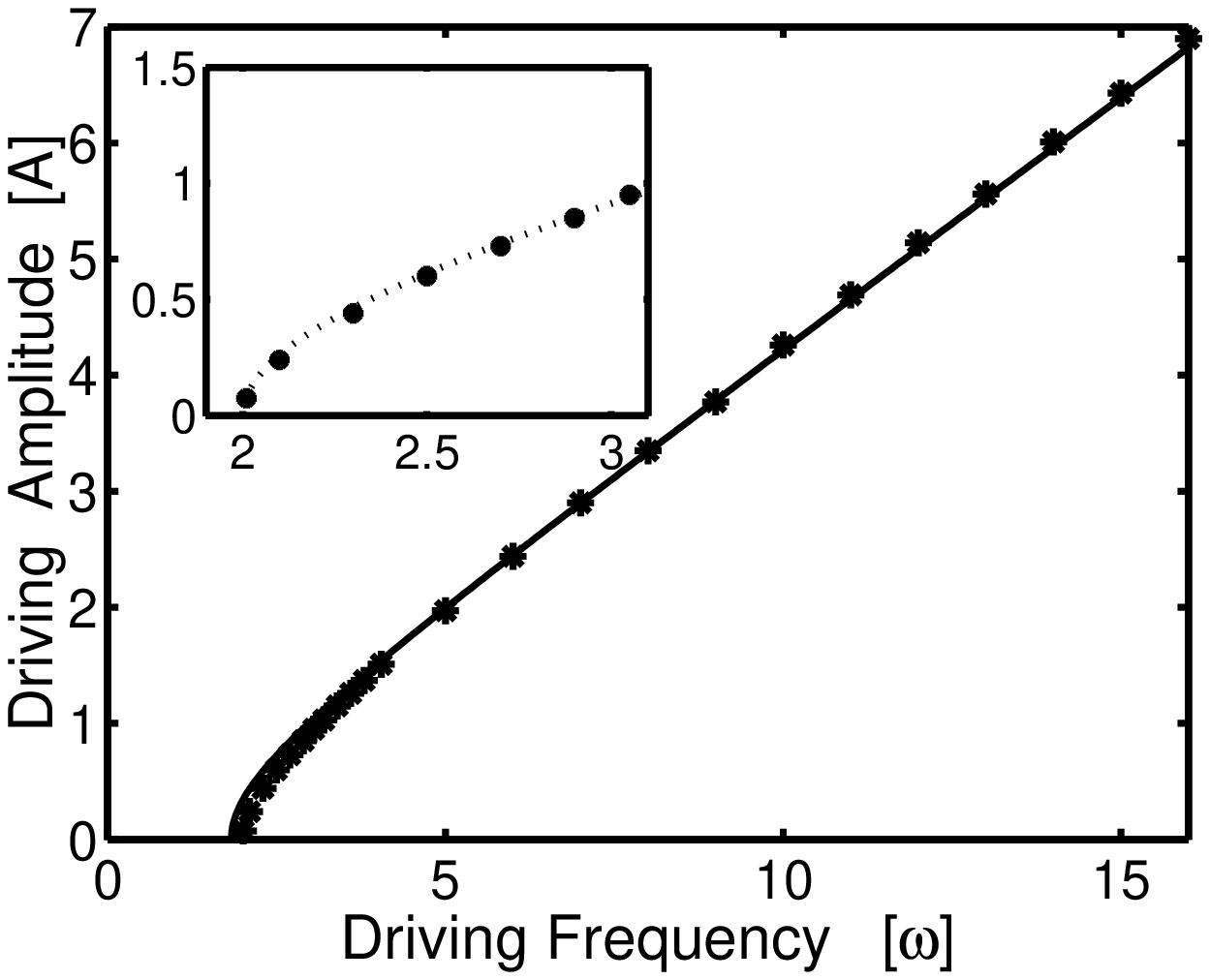}
\caption{Left graph: Snapshot of the local energy
at the transmission threshold $A=0.253 \approx A_{th}$ for $\omega =2.1$
$\gamma=10$. The initial condition is the envelope soliton (\ref{sol})
with $x_0\approx 0$. Right graph: Comparison between analytic estimates and numerical values of threshold amplitudes
vs. the driving frequency.
Main plot: the full dots are the numerical values
of $A_{th}$ and the solid line is a plot of formulae (\ref{6}-\ref{thre}), which are
valid for large amplitudes. The inset shows an enlargement of the small $A_{th}$ region,
in order to illustrate the accuracy of the small-amplitude approximation (\ref{5}) (dotted
line).}
\label{transm}
\end{center}
\end{figure}

The scenario drastically changes at the supratransmission amplitude $A_{th}$. The
chain starts to conduct: a train of {\it travelling} envelope solitons is emitted from
the left boundary (see left panel of Fig.~\ref{transm}). Here we should emphasize that
the envelope soliton solution (\ref{sol}), which is characterized by the $k=\pi$ carrier
wave--number, has a zero group velocity. Thus, transmission cannot be realized by such
envelope solitons.
Instead, transmission starts when the driving frequency resonates with the
frequency of the envelope soliton with carrier wave--number $k=\pi(N-2)/N$, next
to the $\pi$-mode. However, as far as we consider a large number of oscillators ($N=500$),
we can still use expression (\ref{5}) for the $\pi$-mode frequency.

The above envelope soliton solution (\ref{sol})
is valid in the continuum envelope limit, and is therefore less and
less accurate as its amplitude increases. Indeed, if the weakly nonlinear condition
is violated, the width of the envelope soliton becomes comparable with
lattice spacing and, thus, one cannot use the continuum envelope approach.
Fortunately, besides the slowly varying envelope soliton solution (\ref{sol}),
an analytic approximate expression exists for large amplitude static
discrete breather solutions, which is obtained from an exact extended plane wave
solution with ``magic" wave--number $2\pi/3$~\cite{kosprl}
\begin{equation}
u_n=a(-1)^n \cos\left[\omega_B(a)
t\right]\cos\left(\frac{\pi}{3}n\pm x_0\right), \label{breather}
\end{equation}
if $\left|(\pi n/3)\pm x_0\right|<\pi/2$ and $u_n=0$ otherwise.

Here $x_0$ is defined as follows
\begin{equation}
x_0=\mbox{acos}(A/a), \label{initial}
\end{equation}
where $A$ is the driving amplitude. The breather frequency $\omega_B(a)$ depends
on amplitude $a$ as follows
\begin{equation}
\omega_B(a)\simeq 1.03 \frac{\sqrt{3\pi^2(4+9a^2)}}{4K(s)}, \label{6}
\end{equation}
where $K(s)$ is the complete elliptic integral of the first kind with argument
$s=3a/\sqrt{2(9a^2+4)}$ and the factor $1.03$ takes into account a rescaling of
the frequency of the ``tailed" breather~\cite{note} (see also \cite{kos}).
As previously for the case of the envelope soliton solution, we perform a numerical
experiment where we put initially on the lattice the breather solution
of formula (\ref{breather}).
Choosing the plus sign in this expression, we do not observe any significant
transmission of energy inside the chain. Instead, the minus sign
causes the appearance of a moving breather, which travels inside the chain leaving
behind the static breather solution with plus sign. Right graph in Fig.~\ref{2brea} presents this numerical experiment.

The static breather solution (\ref{breather}) ceases to exist
if the driving amplitude exceeds the threshold
$A_{th}$ given by the resonance condition
\begin{equation}
\omega=\omega_B(A_{th}). \label{thre}
\end{equation}
Above this threshold the supratransmission process begins via the emission
of a train of moving breathers from the boundary, exactly as it happens in the
case of small amplitudes. It should be mentioned again that the transmission regime
is established due to moving discrete breathers. It has been remarked~\cite{kosprl}
that discrete breathers are characterized by quantized velocities, while their frequency
is given by the same formula (\ref{6}). This explains why one can use resonance
condition (\ref{thre}) for the static discrete breather solution (\ref{breather})
to define the supratransmission threshold in the large amplitude limit.

\subsubsection{Supratransmission threshold: numerical test}

To check these predictions, we have performed a numerical
determination of $A_{th}$ for several values of $\omega$, starting
the chain at rest. This is accomplished by gradually increasing  $A$
and looking for the minimal value $A_{th}$ for which a sizeable
energy propagates into the bulk of the chain. At early time, the
scenario is qualitatively similar to the one shown in the left graph
of Fig.~\ref{transm}. Later on, the interaction of nonlinear and
quasi-linear modes and their ``scattering"  with the dissipating
right boundary establishes a steady energy flux into the chain.

As seen in the right graph of Fig.~\ref{transm}, formulae
(\ref{thre}) [with definition (\ref{6})] and (\ref{5}) (see the inset) are in excellent
agreement with simulations for large  $A>2$ and small $A\leq 1$ amplitudes,
respectively. The accuracy of the analytical
estimate in formulae (\ref{thre}) and (\ref{5}) is of the order of few percents,
at worst, in the intermediate amplitude range.

For comparison, we have checked that the supratransmission threshold is definitely
not associated with the quasi-harmonic waves with nonlinear
dispersion relation (\ref{2}). If this were the case,  the transmission should
start when the oscillation amplitude reaches the value for which the resonance condition
$\omega=\omega_0(k,A)$ holds. As
$\omega_0(k,A)$ is maximal for $k=\pi$, we can get the expression
for the threshold value from the relation $\omega=\omega_0(\pi,A_{th})$, i.e.
\begin{equation}
\omega^2=4+12A_{th}^2.
\label{3}
\end{equation}
The amplitude values one obtains from Eq.~(\ref{3}) are far away
from the  numerical values and we don't even show them in the right graph of Fig.~\ref{transm}.
This is a further confirmation that supratransmission in the FPU model originates from
direct discrete breather generation as it happens in the cases of discrete sine-Gordon
and nonlinear Klein-Gordon lattices~\cite{leon}.

\section{Conclusions}
\label{conclusions}

In this paper, we have presented a detailed analysis of the
zone-boundary mode modulational instability for the FPU lattice in
both one and higher dimensions. Formulas for the critical amplitude
have been derived analytically and compare very well with numerics
for all system sizes. The study of the process which leads to the
formation of chaotic breathers can be extended to two dimension; the
physical picture is similar to the one-dimensional case.

All results on modulational instability of zone-boundary modes can be
straightforwardly extended to other initial modes and, correspondingly,
instability rates can be derived. This has already been partially done
in Ref.~\cite{Paladin} and compares very well with the numerical
results by Yoshimura~\cite{yoshi2}. This author has recently
reanalyzed the problem~\cite{yoshi3} to determine the growth rates
for generic nonlinearities in the high energy region, obtaining exact
results based on Mathieu's equation.

For many--modes initial excitations, it has been remarked that
instability thresholds depend on relative amplitudes and not only on
the total energy~\cite{chechin2004}. Although this makes the study of
the problem extremely involved, we believe that a detailed study of
some selected group of modes, which play some special role in FPU
dynamics, could be interesting. The method discussed in this paper could
be adapted to treat this problem. Historically, the first study is in
the paper by Bivins, Metropolis and Pasta himself~\cite{bivins}, where
the authors tackle the problem by studying numerically the
instabilities of coupled Mathieu's equations.

In one-dimensional studies, a connection between the average
modulation instability rates and the Lyapunov exponents has been
suggested~\cite{DRT,Paladin}. Recently~\cite{franzozi}, high frequency
exact solutions have been used in the context of a differential
geometric approach~\cite{pettinilyap} to obtain accurate estimates of the
largest Lyapunov exponent. Similar studies could be performed for the
two-dimensional FPU lattice and the corresponding scaling laws with
respect to energy density could be obtained.

Finally, let us point out the generic nature of the results
derived for the driven-damped Anti-FPU scenario. In this connection
further developments towards nonlinear supratransmission and
bistability effects in various physical systems, such as magnetic
thin films~\cite{r1}, Josephson junction arrays~\cite{r2}, quantum
Hall bilayers~\cite{r3}, optical directional couplers~\cite{r4} and
waveguide arrays~\cite{r5,r6} should be expecially mentioned.

\bigskip {\bf Acknowledgement:} We express our gratitude to all our 
collaborators in this field: J.  Barr{\'e}, M. Cl{\'e}ment, T. Cretegny, 
J. Leon, S. Lepri, P. Poggi, A. Torcini. We also thank N. J. Zabusky for useful
exchanges of informations. This work is part of the PRIN contract
{\it Dynamics and thermodynamics of systems with long-range
interactions}. R.Kh. is supported by the Marie-Curie incoming fellowship award 
(MIF2-CT-2006-021328) and USA CRDF Award \# GEP2-2848-TB-06.


\begin{thebibliography}{99}
\bibitem{note} Notice that a simpler approximate expression for the breather frequency
has been proposed in~\cite{kosprl} in the form $\omega_B=\sqrt{3+81A^2/16}$.
We have checked that this expression is also in good agreement with numerical data, but we prefer to use the more accurate form in formula (\ref{6}).
\end{thebibliography}
\end{document}